\documentclass{elsarticle}
\let\today\relax
\makeatletter
\def\ps@pprintTitle{%
    \let\@oddhead\@empty
    \let\@evenhead\@empty
    \def\@oddfoot{\footnotesize\itshape
         {} \hfill\today}%
    \let\@evenfoot\@oddfoot
    }
\makeatother

\biboptions{numbers,sort&compress}       
\usepackage[colorlinks=false]{hyperref}\hypersetup{pdfborder={0 0 0}}
\usepackage{geometry}
\usepackage{amsmath,amsthm,amssymb}
\usepackage{bm}
\usepackage{tcolorbox}
\usepackage{enumitem}
\usepackage{subcaption}
\usepackage{float}
\usepackage{tabularx}
\usepackage{booktabs}
\usepackage{graphicx}
\usepackage{cancel}
\usepackage[numbers]{natbib}
\usepackage[displaymath]{lineno}

\graphicspath{{./figures/}{./results/}{./results/eshelby/}{./results/fracture/}{./results/microstructure/}{./results/microstructure-amr/}}

\definecolor{C0}{HTML}{1F77B4}                                                                                       \definecolor{C1}{HTML}{FF7F0E}                                                                                       \definecolor{C2}{HTML}{2ca02c}                                                                                       \definecolor{C3}{HTML}{d62728}                                                                                       \definecolor{C4}{HTML}{9467bd}   

\begin{document}
\def\interior{\operatorname{int}} 
\def\dom{\operatorname{dom}} 
\begin{frontmatter}

  \title{Massively parallel finite difference elasticity using block-structured adaptive mesh refinement with a geometric multigrid solver}

  \author[uccs]{Brandon Runnels\corref{cor1}}
  \ead{brunnels@uccs.edu}
  \author[au]{Vinamra Agrawal}
  \author[lbl]{Weiqun Zhang}
  \author[lbl]{Ann Almgren}

  \address[uccs]{Department of Mechanical \& Aerospace Engineering, University of Colorado, Colorado Springs, CO, USA}
  \address[au]{Department of Aerospace Engineering, Auburn University, Auburn, AL, USA}
  \address[lbl]{Lawrence Berkeley National Laboratory, Berkeley, CA, USA}
  \cortext[cor1]{Corresponding author}
  
\begin{abstract}
    Computationally solving the equations of elasticity is a key component in many materials science and mechanics simulations.
    Phenomena such as deformation-induced microstructure evolution, microfracture, and microvoid nucleation are examples of applications for which accurate stress and strain fields are required. 
    A characteristic feature of these simulations is that the problem domain is simple (typically a rectilinear representative volume element (RVE)), but the evolution of internal topological features is extremely complex. 
    Traditionally, the finite element method (FEM) is used for elasticity calculations; FEM is nearly ubiquituous due to (1) its ability to handle meshes of complex geometry using isoparametric elements, and (2) the weak formulation which eschews the need for computation of second derivatives.
    However, variable topology problems (e.g. microstructure evolution) require either remeshing, or adaptive mesh refinement (AMR) - both of which can cause extensive overhead and limited scaling.
    Block-structured AMR (BSAMR) is a method for adaptive mesh refinement that exhibits good scaling and is well-suited for many problems in materials science.
    Here, it is shown that the equations of elasticity can be efficiently solved using BSAMR using the finite difference method.
    The boundary operator method is used to treat different types of boundary conditions, and the ``reflux-free'' method is introduced to efficiently and easily treat the coarse-fine boundaries that arise in BSAMR.
    Examples are presented that demonstrate the use of this method in a variety of cases relevant to materials science: Eshelby inclusions, fracture, and microstructure evolution. 
    Reasonable scaling is demonstrated up to $\sim$4000 processors with tens of millions of grid points, and good AMR efficiency is observed.
\end{abstract}
\end{frontmatter}

\section{Background and Motivation}

Materials are well-known to exhibit drastically different properties on different scales \cite{meyers1994dynamic}.
Polycrystalline materials such as metals typically behave isotropically on the macroscale, but exhibit a non-trivial stress fields at the microscale (even under uniform loading) resulting from their complex grain structure.
The sub-grain response, in turn, is determined by the crystalline structure and the existence of defects (such as dislocations) at the atomic scale.
Therefore, to obtain material response informed by first-principles calculations, it is necessary to bridge spatial and temporal scales.
A common way to link scales is by simulating complex behavior that takes place on the mesoscale within representative volume elements (RVEs).
RVE simulations incorporate atomistically-informed physics (such as grain boundary and plasticity models \cite{keshavarz2013multi,keshavarz2015hierarchical,jafari2017constitutive}) to provide continuum-level mechanical data (such as elastic moduli or yield stress).
RVE simulation domains can exist on scales ranging from nanometers to centimeters (depending on the application), but are common in that the simulation domain is rectilinear and periodic.
The features of interest are neither the geometry of the domain nor the boundary conditions, but rather the material discontinuities within the RVE and the resulting heterogeneous response.

A common example of such RVE simulations is the evolution of microstructure in metals, in which the grain boundaries (GB) move to reduce their energy and the elastic energy of the material.
This presents the problem of generating a suitable discretization that adapts to the ever-changing microstructure in the RVE.
One way of addressing this problem is though the use of diffuse interface methods such as the multi-phase field (MPF) method for microstructure evolution \cite{vedantam2006efficient,schmitz2010phase}.
In this method, rather than mesh the microstructure explicitly, a uniform grid is used to resolve the transition region between grains. 
This method is highly advantageous in that it allows for fully arbitrary microstructure evolution and is not hindered by topological transitions such as boundary splitting/merging.
However, it can also be computationally expensive due to the excessive number of trivial grid points located within the interior of each grain.
To resolve this problem, adaptive mesh refinement (AMR) is used to selectively refine the regions of interest across GBs, while reducing the number of trivial points in the grain interiors.
The adaptivity of the method allows the mesh to be freely driven by the evolution of the problem.

Many flavors and varieties of AMR are available \cite{plewa2005adaptive}.
Block-structured AMR (BSAMR) is an AMR method in which the mesh is organized into ``levels'' such that each level is comprised of grid cells that are all the same size \cite{dubey2014survey}.
Each level is treated independently, and then coarse levels are updated with information from fine levels using averaging.
BSAMR is highly scalable and is well-suited for massive parallelism, compared to other methods that exhibit a prohibitive amount of communication overhead as the mesh size increases.
This method is a prime candidate for use in RVE based problems where domain geometry often rectilinear.

The Finite Element Method (FEM) is ubiquitous in computational science and engineering, particularly for solving elasticity problems. 
FEM algorithms have evolved for adaptive grids and extreme scalability to study a wide range of complex solid mechanics problems with intricate geometries and boundary conditions.
However, for RVE based problems with evolving microstructure and evolving boundaries, the advantages of FEM are less relevant.
Here, we propose that the use of the Finite Difference Method (FDM) is advantageous for some of the proposed problems in elasticity, and propose an efficient method of solving the elastic FDM problem using BSAMR.
We demonstrate the use of FDM in BSAMR, and develop a new approach for accurately treating the boundary between AMR levels (the ``coarse-fine boundary'').
While parallel AMR Finite Element Method (FEM) implementations have been implemented \cite{neiva2019scalable}, scalability is ultimately limited by the choice of AMR that is used.
Finite difference (FD), on the other hand, lends itself easily to a BSAMR implementation.

Relatively little work has been done to develop the FDM for elasticity.
While there has been some work to solve for stresses and displacements using finite difference with a displacement potential function \cite{morales2013numerical,harangus2014finite}, it has not been applied on a large scale.
In this work, we discuss the FDM for finite difference, introduce a BSAMR implementation, and present several numerical results.
This paper is organized in the following way.
Section~\ref{sec:theory_and_computation} reviews the formulation of elasticity and describes the FDM implementation.
Section~\ref{sec:implementation} describes the computational framework (``Alamo''), the block-structured adaptive mesh refinement, and the multigrid solver. 
Section~\ref{sec:examples} demonstrates the use of the code applied to several classic elastic problems, as well as massively parallel microstructure simulations.
Performance and scaling data are discussed in section~\ref{sec:performance}, and we conclude with conclusions and acknowledgments in section~\ref{sec:conclusions}.

\section{Theory and Computation}\label{sec:theory_and_computation}

In this section we review linear elasticity in the strong form only, and the practical challenges involved in constructing a finite-difference implementation.
Next, we discuss block-structured AMR and the geometric multigrid method for solving linear system.
Finally, we introduce the ``reflux-free'' method for treating multi-level solutions at coarse-fine boundaries.

\subsection{Finite difference method for elasticity}

Let $\Omega\subset\mathbb{R}^d$ be the $d$-dimensional problem domain with boundary $\partial\Omega$.
The subset $\partial_1\Omega\subset\partial\Omega$ is the portion of the boundary subjected to {\it displacement} boundary conditions (also referred to as Dirichlet or essential BCs.)
The subset $\partial_2\Omega\subset\partial\Omega$ is the boundary subjected to {\it traction} boundary conditions (also referred to as natural BCs.)
The subset $\partial_3\Omega\subset\partial\Omega$ is the part of the boundary subjected to Neumann boundary conditions.
We note that there is an important distinction between traction ($\partial_2\Omega$) and Neumann ($\partial_3\Omega$) conditions and they are not interchangeable.

The following are the governing equations for static linear elasticity in Cartesian coordinates:

\begin{subequations}
\begin{align}
  &\frac{\partial}{\partial x_j}\Big[ \mathbb{C}_{ijkl}(\bm{x})\Big(\frac{\partial u_k}{\partial x_l} - \varepsilon^0_{kl}\Big)\Big] = - b_i(\bm{x})  && \forall \bm{x}\in\operatorname{int}(\Omega)\label{eq:elastic_int}\\
  &u_i = u^0_i(\bm{x}) &&\forall \bm{x}\in\partial_1\Omega \label{eq:elastic_b1}\\
  &\mathbb{C}_{ijkl}(\bm{x})\Big(\frac{\partial u_k}{\partial x_l} - \varepsilon^0_{kl}\Big)\,n_j(\bm{x}) = t^0_i(\bm{x}) && \forall \bm{x}\in\partial_2\Omega \label{eq:elastic_b2}\\
  &\frac{\partial u_i}{\partial x_j}\,n_j(\bm{x}) = \delta u^0_i(\bm{x}) && \forall \bm{x}\in\partial_3\Omega. \label{eq:elastic_b3}
\end{align}
\end{subequations}
The tensor $\bm{\varepsilon}^0$ is an eigenstrain, representing e.g. plastic slip or thermal strain, and is $0$ in the linear elastic case.
$\mathbb{C}$ is the fourth-order, spatially varying elasticity tensor, $\bm{b}$ is the body force vector field, and $\bm{u}^0, \bm{t}^0, \delta\bm{u}^0$, are the prescribed displacements, tractions, and displacement gradients, respectively.

In the present finite difference formulation, boundary conditions are treated as boundary operators.
Equation~\ref{eq:elastic_int} is represented by an operator $D:\operatorname{int}(\Omega)\to\mathbb{R}^d$ acting on displacements, for which the negative body force is the right hand side.
Similarly, $\partial_1D$, $\partial_2D$, and $\partial_3D$ are operators acting on displacement vectors in the boundary, for which the right hand sides are prescribed displacement, traction, and displacement gradient, respectively.

\begin{figure}[t]
  \centering
  \includegraphics{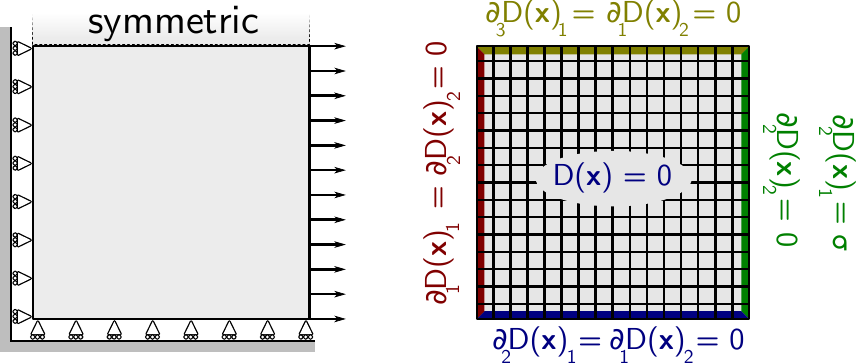}
  \caption{Example of a partitioned boundary operator}
  \label{fig:partitionedoperators}
\end{figure}

Figure~\ref{fig:partitionedoperators} illustrates the case of a loaded block with the use of partitioned operators as surrogates for boundary conditions.
The partitioned operator is then implemented in such a way that the linear solver can be entirely agnostic to boundary conditions, with boundary condition information passed in as components of the right hand side. 

\subsubsection{Affine elasticity and material discontinuities}

Material RVE problems often involve dealing with material discontinuities such as grain boundaries, material interfaces, pores and microcracks. 
One way to treat discontinuities (such as shock waves in fluids or fluid-structure interaction) is by explicitly tracking the sharp interface; such approaches include front-tracking methods \cite{chern1986front,terashima2009front,agrawal2018impact}, level set methods \cite{sethian2001evolution,gibou2018review}, and immersed boundary methods \cite{beyer1992computational,li1994immersed,roma1999adaptive,peskin2002immersed}.
On the other hand, diffuse interface methods require mollification of the discontinuity to ensure smoothness of derivatives, while front tracking algorithms move the mesh such that interfaces lie on mesh boundaries. 
Diffuse interface methods are a natural choice for most of the applications proposed here, and work naturally with AMR to provide selective refinement at the diffused boundary.
Diffuse interfaces are also very convenient for linear elasticity, as strain compatibility and stress jump conditions are satisfied automatically without explicit treatment.
Here, to tackle material discontinuities, we employ the diffused interface approach.
We simply require that $\mathbb{C}_{ijkl}$ and $\bm{\varepsilon}$ be smooth functions of $\bm{x}$.
When the need arises to model a discontinuity, a transition (or ``diffuse'') region will be introduced over which the values will be continuously varied.
This requires that the mesh be highly refined near the transition region, which is a simple matter with AMR.
Indeed, for many applications, other problem constraints already required high resolution at the boundary, so this does not incur any additional computational cost.

For a problem with spatially varying elastic properties, the operator (\ref{eq:elastic_int}) is split using the product rule
\begin{align}
  D(\bm{u})_i = \mathbb{C}_{ijkl,j}(u_{k,l} - \varepsilon^0_{kl}) + \mathbb{C}_{ijkl}(u_{k,lj}-\varepsilon^0_{kl,j}),
\end{align}
avoiding the need to create a surrogate stress field.
Operator overloading is used extensively to efficiently compute the modulus derivative $\mathbb{C}_{ijkl,j}$ in a portable way.
High symmetry tensors (such as those corresponding to isotropic or cubic materials) store only the reduced number of constants, which not only saves space but also significantly speeds up the computation time for the derivative.

For problems with eigenstrain, the operator is decomposed into homogeneous and inhomogeneous components.
The homogeneous operator is readily computed to be

\begin{align}
  D_H(\bm{u}) = D(\bm{u}) - D(\bm{0})
\end{align}

Thus the solution to $D(\bm{u})=-\bm{b}$ is given simply by the solution to 

\begin{align}
  D_H(\bm{u}) = -(D(\bm{0}) + \bm{b}),
\end{align}

\noindent that is, the inhomogeneous part of the operator is transferred to the right hand side, and acts as a source term in the solution, provided that the eigenstrain field $\bm{\varepsilon}^0$ is $C_1$ continuous.

\begin{figure*}
  \includegraphics[width=\linewidth]{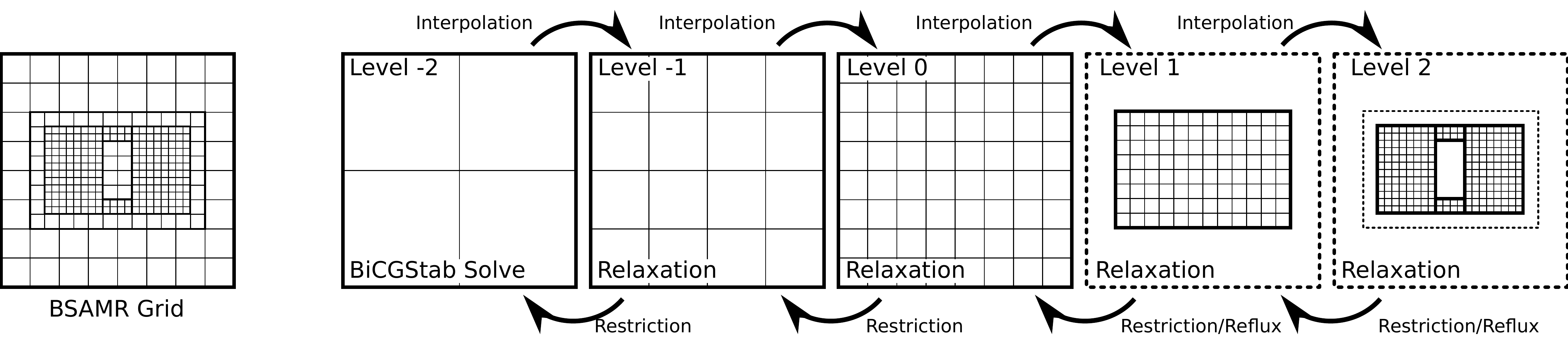}
  \caption{Structure of the multi-level/multi-grid solver, illustrating the continuity between multigrid (level $<0$) and levels of refinement (level $>0$).}
  \label{fig:hierarchy}
\end{figure*}

\section{Implementation}\label{sec:implementation}

All results were generated using the implementation in Alamo, a C++ code developed by the authors, which is built on the AMReX library.
Zhang {\it et al} \cite{zhang2019amrex} is a general reference for this section.
For clarity, we define nomenclature in the following table:

\begin{figure}[h]
  \begin{tabularx}{\linewidth}{l X}
    \toprule 
    $d$ & Spatial dimension \\
    $i,j,k,l$ & Indices corresponding to coordinates \\
    $m,n$ & Indices corresponding to ``fine'' and ``coarse'' AMR levels \\
    $\Omega_{m}$ & Domain for refinement level $m$ \\
    $\partial\Omega^{CF}_{m}$ & Coarse-fine boundary (not domain boundary)  \\
    $\bm{r}_n$ & Right hand side on level $n$ \\
    $D_{n}$ & Linear operator on level $n$  \\
    $S^D_{n}$ & Relaxation operator on level $n$  \\
    $I_{nm}$ & Interpolation function \\
    $R_{mn}$ & Restriction function  \\
    \bottomrule
  \end{tabularx}
  \caption{Notation}
  \label{tab:notation}
\end{figure}

\subsection{Geometric multigrid solver}

Many algorithms for multiphysics PDE-based applications require the iterative solution of large linear systems arising from either discretization of elliptic equations or implicit treatment of parabolic equations.
Multigrid algorithms are often the method of choice for solving these systems.
AMR simulations with subcycling in time typically require linear solves across the grids at a single AMR level only; algorithms without subcycling typically require solves across all the levels in the hierarchy.
For the rest of this paper we focus on the performance of multigrid on multiple levels of an AMR hierarchy, using either V-cycles or F-cycles (see, e.g., \cite{briggs2000multigrid}, for more detail).

The multigrid method can be naturally combined with block-structured AMR by viewing the refined levels as an extension of the coarse/fine multigrid level sequence (Figure~\ref{fig:hierarchy}).
In this work, we refer to the initial, unrefined grid as ``Level 0,'' the first level of refinement as ``Level 1,'' and so on.
The first multigrid coarsening level is ``Level -1'' the second is ``Level -2,'' etc.
A full discussion of multigrid implemented with block-structured AMR is discussed in \cite{almgren1998conservative}.

In the present work, the multigrid solver with FDM is used with F-cycles only. 
For each smoothing operation, two sequential Jacobi iterations are performed.
The solver runs a total of four smoothing operations per level, for a total of eight Jacobi iterations.
It should be noted that in order for the operator (stiffness matrix) to be consistently refined/coarsened, there must be operator equivalence at each level.
The only elements that satisfy this equivalence are bilinear (2D) and trilinar (3D) rectangular elements; however, these are well-known to exhibit shear-locking and other undesirable effects \cite{stolarski1983shear} in FEM.
On the other hand, the FDM is easily applied and requires approximately the same number of calculations as FEM.

To manage mesh refinement and coarsening, individual cells or nodes are tagged as needing refinement.
The criteria for tagging is different for each test case, but is generally based on the derivative of an indicator field specific to the problem \cite{li2010comparison}.
The list of tagged cells are then used by AMReX to iteratively regrid each level until either the refinement criteria are met or the maximum allowable refinement level is reached.

\subsection{Reflux-free BSAMR extension}

A main contribution of this work is a novel method for restriction between coarse and fine AMR levels.
We proceed in this section by formally defining the necessary properties for relaxation, interpolation, and prolongation.
Some of the notation is enumerated in Table~\ref{tab:notation}. 

\paragraph{Relaxation}
The relaxation operator $S^D_n:[C(\Omega_n)]^2\to C(\Omega_n)$ is defined such that
\begin{align}
  D_n\Big[\lim_{q\to\infty} (S^D_n)^q (\bm{\phi_n})\Big] = \bm{r}_n \ \ \ \text{ on } \interior(\Omega_n)
\end{align}
that is, the relaxation operator converges to the exact solution with $\partial^{CF}\Omega_m$ acting as part of $\partial_1\Omega_m$ (Dirichlet boundary).

\paragraph{Interpolation}
The interpolation operator $I_{nm}:C(\Omega_m\cap\Omega_n) \to C(\Omega_{m})$ projects the solution from the coarse level to the fine level.
We require that $I_{nm}$ satisfy the following properties:
\begin{enumerate}
\item
  Translation symmetry: $I_{nm}(\phi \,\circ\, \bm{t}) = I_{nm}(\phi) \, \circ \, \bm{t}$ where $\bm{t}:\Omega_m\to\Omega_m$ is an admissible translation operation.
  This means that the interpolation operator should be the same everywhere in the pre-image of $I_{nm}$.
  (Admissible = does not translate outside of pre-image of $I_{nm}$.)
\item 
  Injectivity: in the special case of BSAMR refinement, level $m$ has higher resolution than level $n$.
\end{enumerate}

\paragraph{Restriction}
The restriction operator $R_{mn}:C(\interior\Omega_{m})\to C(\interior\Omega_m\cap\Omega_n)$ projects the solution from the fine level to the coarse level.
$R_{mn}$ must satisfy the following properties
\begin{enumerate}
\item
  Translation symmetry: complementary to interpolation operator.
\item 
  Surjectivity: in the special case of BSAMR refinement, level $n$ has lower resolution than level $m$.
\item
  Nonlocality: $(R_{mn}\circ\,\phi^n)^{-1}(\bm{x}_n) = B_\varepsilon(\bm{x}_m)$ where $B_{\varepsilon}$ is an $\varepsilon$-neighborhood of $\bm{x}_m$.
  That is, the value of the restricted function at the coarse point $\bm{x}_n$ requires information at {\it and around} fine-level point $\bm{x}_m$.
\end{enumerate}

\par We see that together, properties of translation symmetry and nonlocality require additional restrictions be placed on the domain of $R_{mn}$.
Consider a point $\bm{x}_m\in\partial^{CF}\Omega_m$: nonlocality requires dependence on $B_{\varepsilon}(\bm{x})$; therefore, this becomes boundary-dependent.
To preserve translation symmetry, we make the domain of the restriction operator $\dom(R_{mn})=C(\interior_\varepsilon\Omega_{m})$, where the ``$\varepsilon$-interior'' is

\begin{align}
  \interior_\varepsilon\Omega_{m} = \{\bm{x}_m\in\Omega_m:B_{\varepsilon}(\bm{x}_m)\subset\Omega_m).
\end{align}

We are now left with the problem of restricting points on the coarse/fine boundary $\partial^{CF}(\Omega_m)$.
Improper treatment of coarse/fine boundary can result in mismatched operator, leading to artificial forces and poor convergence.
One possibility is to augment the restriction operation with a special operator with mixed stencil that uses information about $D$ to both smooth and restrict $\bm{\phi}_m$.
This was proposed by Almgren {\it et al} \cite{almgren1998conservative} as a ``reflux'' operation, and has been demonstrated to work successfully in many applications \cite{sverdrup2018highly,ream2019numerical,vay2018warp,nonaka2019amrex,fullmer2019benchmarking}.
However, when the linear operator $D$ is complicated, the practical implementation of such a reflux operation may become prohibitively difficult to implement.
Furthermore, it lacks generality as it requires operator-specific information for restriction. 

\begin{figure}[t]
  \centering
  \includegraphics[width=0.5\linewidth]{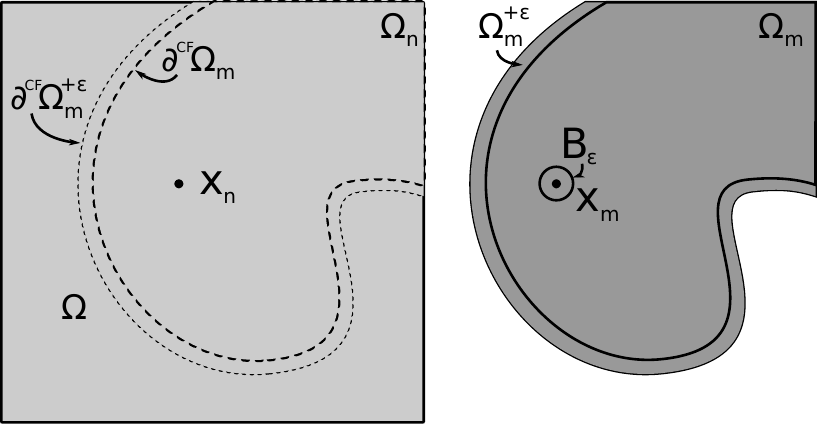}
  \caption{The fine level ($\Omega_m$) must be grown by $\varepsilon$ in order for the restriction operation to maintain translation symmetry.}
  \label{fig:domains}
\end{figure}

We propose an alternative modification.
Let us define the $\varepsilon-$neighborhood of $\Omega_m$ (Figure~\ref{fig:domains})

\begin{align}
  \Omega^{+\varepsilon}_m = \{\bm{x}:\exists \bm{x}_m\in\Omega_m\text{ such that } \bm{x}\in B_{\varepsilon}(\bm{x}_m)\}.
\end{align}
Then the pre-image of $R_{mn}\circ\bm{\phi}_n$ is exactly $\interior(\Omega_m^{+\varepsilon})$. 
Numerically, the interpolation operator is implemented as a basic bilinear (2D) or trilinear (3D) interpolation from the coarse level to the fine level.
A standard restriction matrix (with powers of 2) is used for the restriction operator.
If the restriction operator acts on each of the $N$-nearest neighbors, then the fine AMR level is grown to include exactly $N+1$ ``ghost nodes'' (corresponding to the discretized $\partial^{+\varepsilon}\Omega$).
(In general, $N=1$ for all cases considered here.)
A detailed description of the resulting effective stencil is worked out in \ref{sec:coarsefinestencil}

In a numerical multigrid scheme, the reflux-free method works in the interpolation-smooth-restriction cycle in the following way:
For each cycle, the interpolation operator updates all non-boundary fine level nodes with updated values from the coarse level.
The fine level is then relaxed, with the outer layer of ghost nodes acting as a Dirichlet boundary with values set by the coarse level.
However, all other nodes have been properly updated, including those nodes on the coarse/fine boundary.
Therefore, when the solution is restricted from the fine level to the coarse, the pre-image of the restriction operator on the coarse grid is the entirely updated solution.
This implementation also preserves the necessary properties of translational symmetry and nonlocality of restriction operator.
Finally, this eliminates the need for special treatment, such as a mixed stencil or operator dependent reflux operation, on the coarse/fine boundary.
Since $\partial^{CF}_m$ is, by definition, on the interior of the domain, the $\epsilon-$neighborhood is then filled using the interpolation operator.

\section{Examples}\label{sec:examples}

In this section we present a collection of numerical examples that show the versatility of the method, provide verification against known solutions, and demonstrate scalability and performance.
Each section provides brief background on the methods involved in the calculation, but in each case, the literature should be consulted for a full discussion.

\subsection{Eshelby inclusion}

The Eshelby inclusion for an ellipsoid is a standard problem in linear elasticity that was originally introduced (and solved) by Eshelby \cite{eshelby1957determination}.
The problem is generally described as follows: consider an ellipsoidal region in a homogeneous, infinite, stress-free solid.
Let the ellipsoid be removed from the solid and replaced by an ellipsoid that has been permanently deformed by a strain $\bm{\varepsilon}^{eshelby}$, called an {\it eigenstrain}.
Once the body is allowed to relax, the mismatch induces a residual strain in both the inclusion and the matrix.  

Eshelby introduced an analytical solution for this problem and derived the ``Eshelby tensor'' for the ellipsoidal inclusion \cite{eshelby1959elastic}.
However it was not until recently that the displacement field (and subsequent strain and stress fields) for an arbitrary ellipsoidal inclusion was derived \cite{jin2016explicit}. 
This ellipsoid inclusion problem is a useful test case for the present elasticity model, as it enables verification of the use of diffuse material discontinuities as well as far-field stresses.

\begin{figure}[t]
  \centering \includegraphics[width=0.5\linewidth]{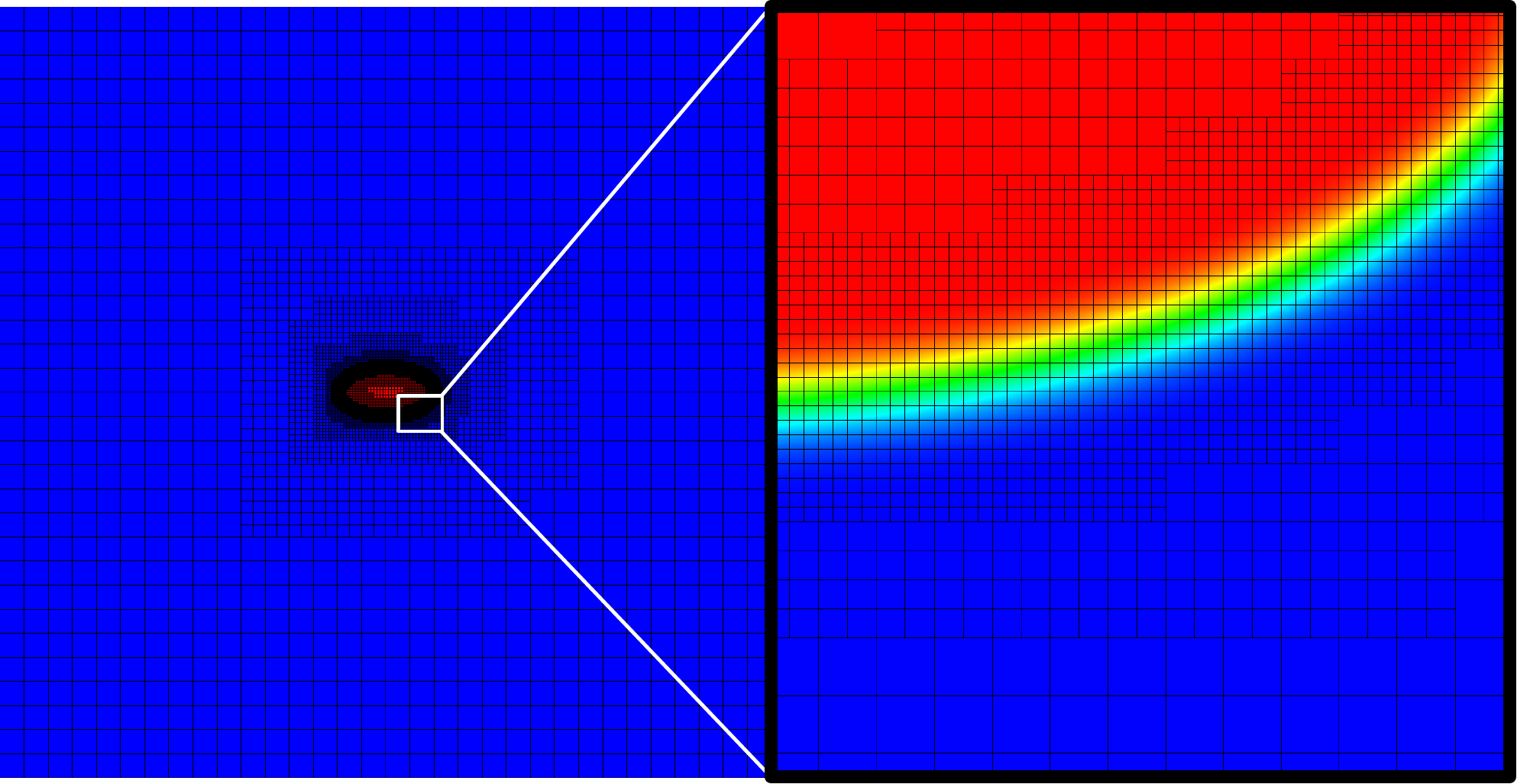}
  \caption{
    (Eshelby inclusion)
    Slice of 3D grid with 6 total AMR levels.
    Blue (exterior) indicates $\bm{\varepsilon}^0=\bm{0}$; red (interior) indicates $\bm{\varepsilon}=\bm{\varepsilon}^{eshelby}$.
  }
  \label{fig:eta_etamesh}
\end{figure}

The problem is constructed in a 8x8x8 (arbitrary length units) domain with fixed-displacement boundary conditions.
The large domain approximates the ``infinite body'' used in the classic Eshelby problem.
The base-level grid is is $32^3$, and there are 5 levels of refinement for 6 total AMR levels (Figure~\ref{fig:eta_etamesh}).
The width of the diffuse boundary is 0.1 to ensure adequate grid coverage.
The radii for the ellipse in the $x_1,x_2,x_3$ directions are $a_1=1.0, a_2 = 0.75, a_3 = 0.5$, the elastic moduli are $E=210,\nu=0.3$, and the eigenstrain is $\bm{\varepsilon}^{eshelby} = \operatorname{diag}(1E-3,1E-3,1E-3)$ and zero elsewhere. 
(All values taken from \cite{jin2016explicit}.)

The sharp-boundary ellipsoid is determined by the indicator function 

\begin{align}
  \eta(\bm{x}) = \begin{cases} 1 & \bm{x}^TA\bm{x}>1 \\ 0 & \text{else}\end{cases},
\end{align}
(with $A_{ij}=(\varepsilon^{eshelby}_{ij})^{-2}$).
The diffuse indicator function $\eta_\epsilon$ is constructed in the following way:

\begin{align}
  \eta_\varepsilon(\bm{x})  = \frac{1}{2}\Big(1 - \operatorname{erf}\Big(\frac{\bm{x}^TA\bm{x}}{2\,\epsilon\,|A{\bm{x}}|}\Big)\Big).
  \label{eq:ErrorFunctionMollification}
\end{align}
Since $2A\bm{x}$ is the slope of $\eta$, dividing the argument of the error function by the magnitude of its slope ensures a generally uniform diffuse region.
The eigenstrain field is then given by

\begin{align}
  \bm{\varepsilon}(\bm{x}) = \eta(\bm{x}) \, \bm{\varepsilon}^{eshelby}.
\end{align}

To solve the problem, an AMR mesh is used with a total of 1,010,240 grid points, and the finest level consists of 667,712 nodes concentrated over 0.062\% of the domain volume.
The criteria
\begin{align}
  |\nabla \eta(\bm{x})|\,|\Delta \bm{x}| > 0.01
\end{align}
was used to tag cells for regridding ($\Delta\bm{x}$ is the vector of grid spacings).
As the focus of this test was validation, the parameters of the numerical solution are not optimized for performance; for instance, it does not take advantage of symmetry.
However, the MLMG solver completed in approximately 2 minutes on a laptop running with 6 MPI processes.
The convergence factor was 0.1 for the first few iterations before reaching a constant factor of 0.5 after 2-3 iterations.

\begin{figure}[t]
  \centering\includegraphics[width=0.5\linewidth]{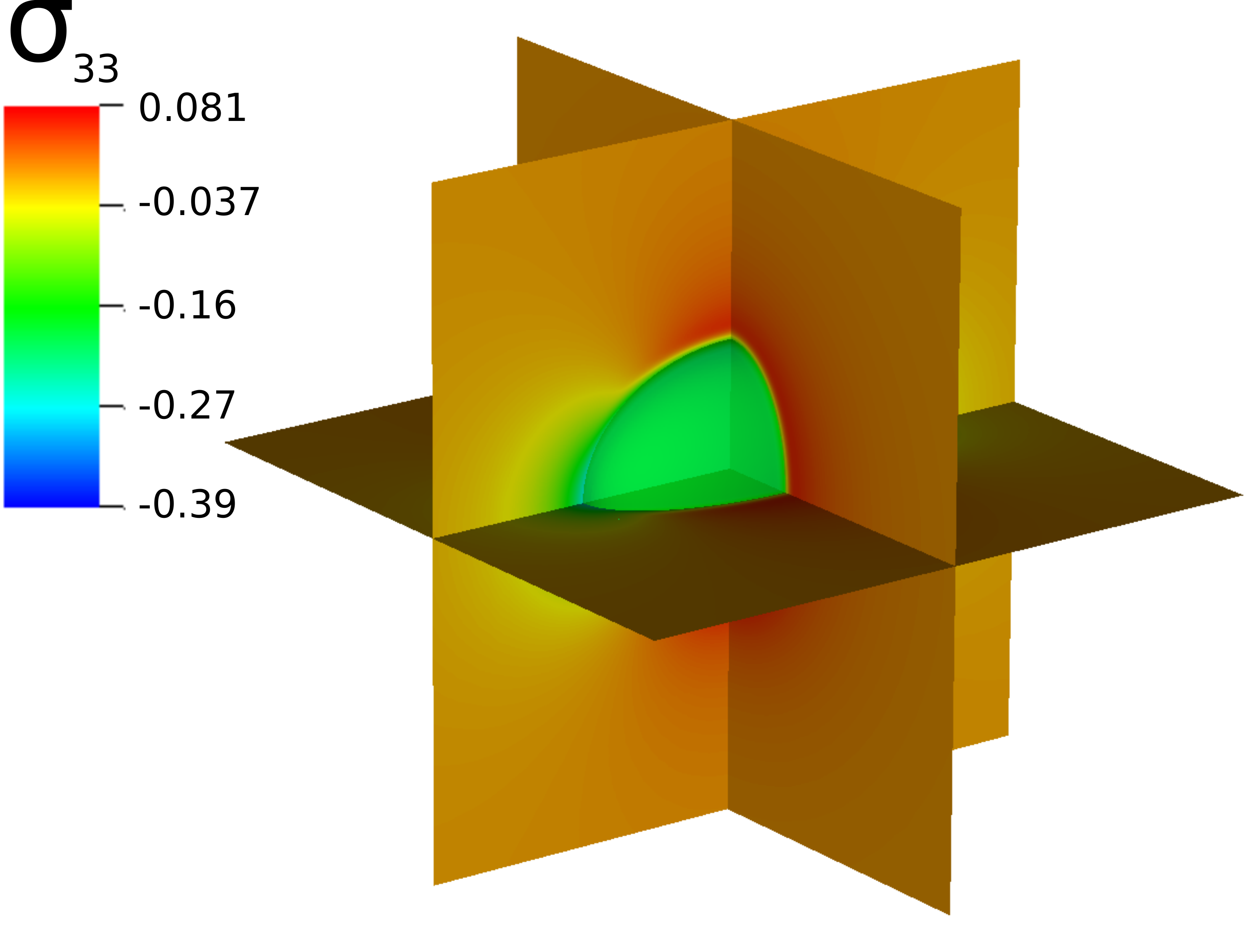}
  \caption{(Eshelby inclusion) Slice plot of $\sigma_{33}$ (green ellipse indicates the boundary of the inclusion}
  \label{fig:eshelby_stress_3d}
\end{figure}

Stresses and displacements are determined for the entire region (Figure~\ref{fig:eshelby_stress_3d}).
It is determined that all displacements, strains, and stresses drop off to zero at a substantial distance from the domain boundary.

\begin{figure}[H]
  \begin{subfigure}{0.5\linewidth}
    \includegraphics[width=\linewidth]{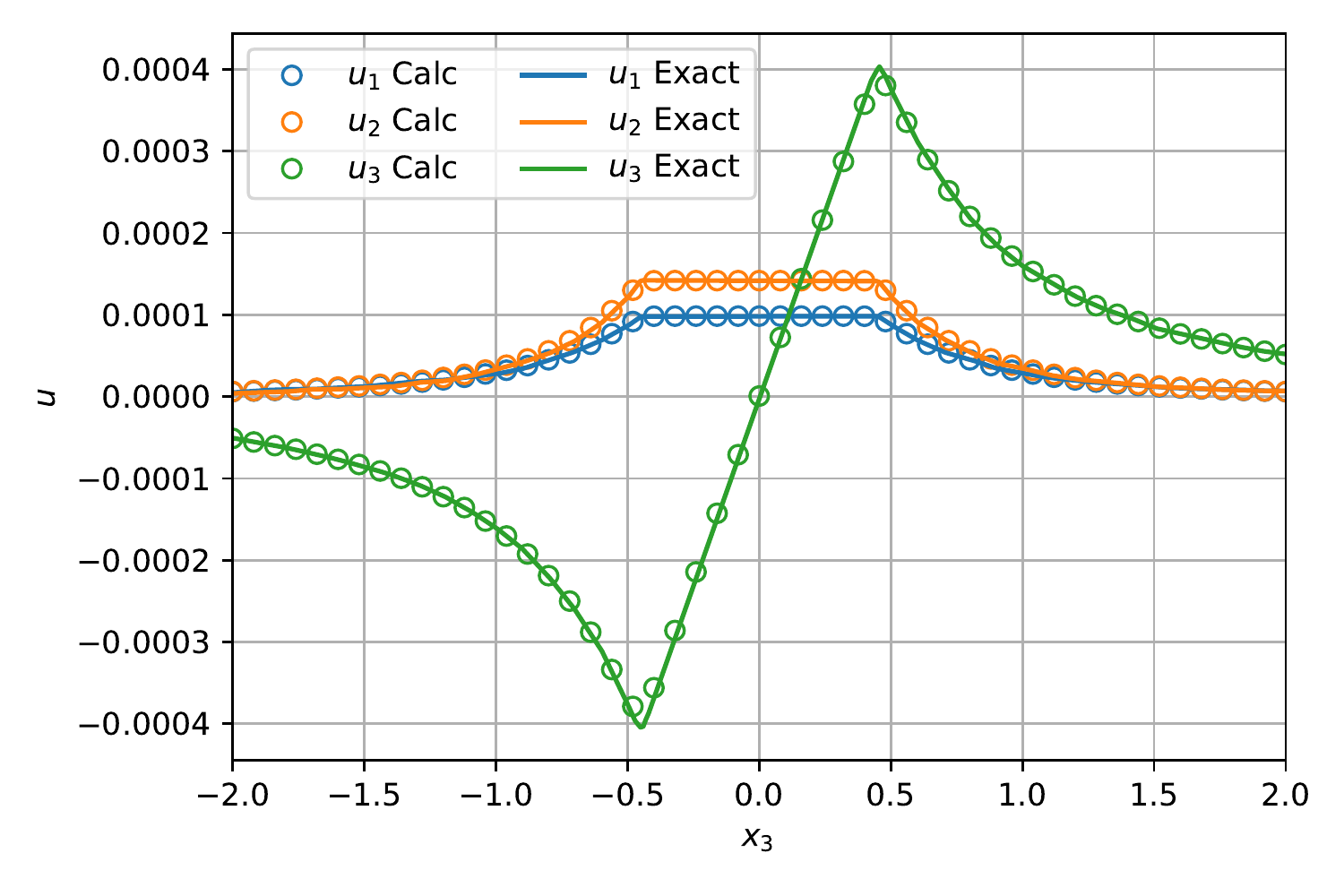}
    \caption{Displacements vs analytical solution}
    \label{fig:Eshelby_displacement}
  \end{subfigure}
  \begin{subfigure}{0.5\linewidth}
    \includegraphics[width=\linewidth]{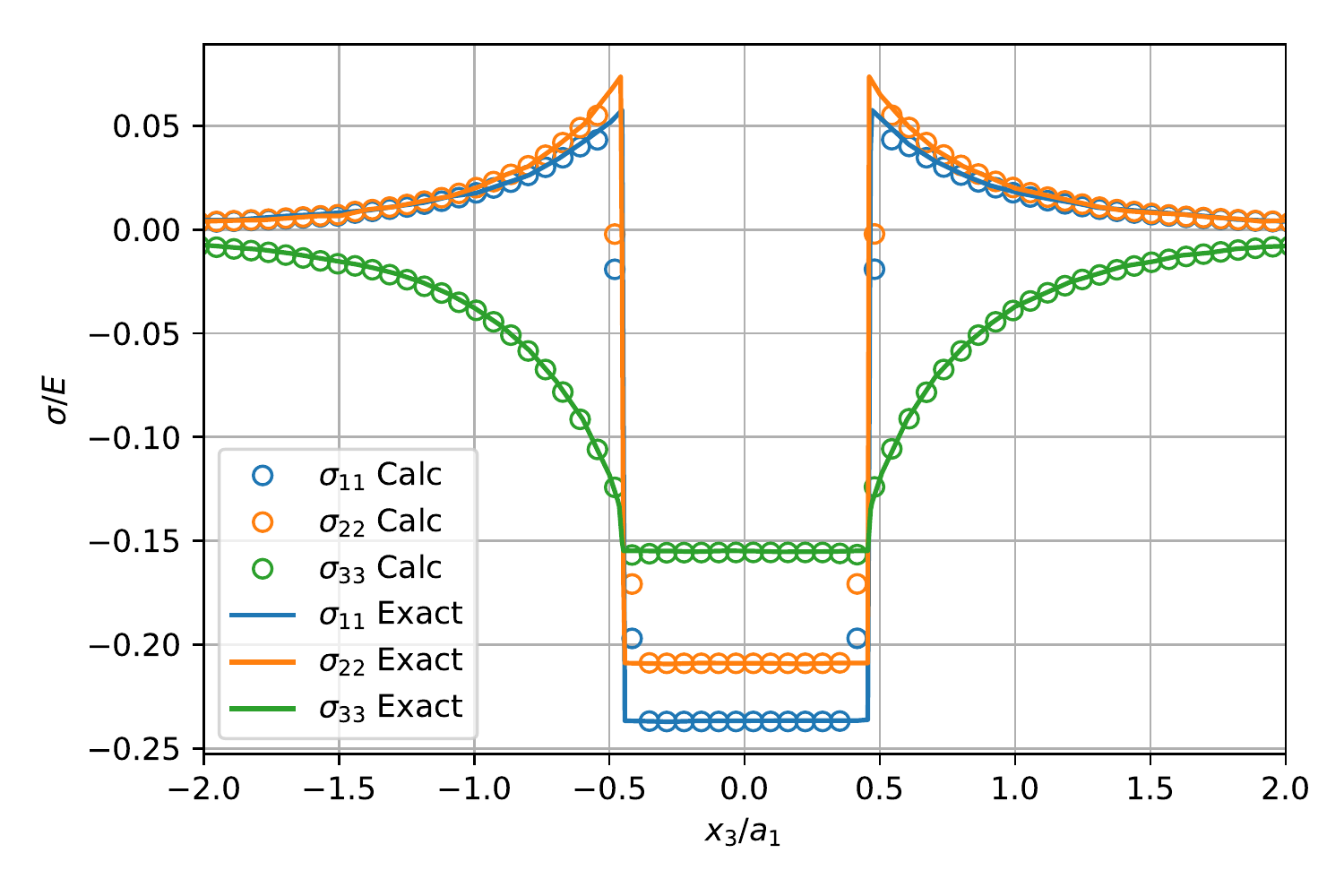}
    \caption{Stresses vs. analytical solution }
    \label{fig:Eshelby_sigma_normal}
  \end{subfigure}
  \caption{(Eshelby inclusion) Comparison of displacements and normal stresses to analytical solution as derived in \cite{jin2016explicit}}
\end{figure}

To validate the solution, displacements, strains, and stresses are compared to exact values in Jin {\it et al} \cite{jin2016explicit}.
The values are determined along the line between the points $[0.25,0.25,-4]$ and $[0.25,0.25,4]$.
Comparison is included here for displacements (Figure~\ref{fig:Eshelby_displacement}) and normal stresses (Figure~\ref{fig:Eshelby_sigma_normal}).


For the displacements, a relative error of 1.6\%, 2.8\%, and 1.6\% is observed for $u_1$, $u_2$, and $u_3$, respectively.
An examination of the data indicated that nearly all of the error was concentrated at the diffuse boundary; however, it was observed that this error decreased with respect to the diffuse boundary width.
The stress is shown to be discontinuous, but the diffuse boundary smooths out the discontinuity and allows the stress to vary smoothly between the interior and exterior of the inclusion.
There is no substantial error in the solution away from the diffuse boundary.

\subsection{Fracture}

Catastrophic and incremental failure in solid materials is often mediated by the nucleation and propagation of fractures.
Crack propagation in brittle and ductile materials has been a topic of study for decades.
With recent advancements in additive manufacturing methods, modeling fatigue-induced crack propagation in additively manufactured metals is particularly important \cite{torries2018overview}. 

One of the many prominent techniques of modeling fracture is cohesive zone modeling.
This technique involves defining a traction-separation relationship across fracture surfaces \cite{park2011cohesive}.
Some cohesive zone models introduce potential functions to define the nonlinear traction-separation relationship. 
Cohesive zone models have been successfully used to capture effects of crack tip radius and plasticity effects near the crack tip \cite{jin2006comparison}.
However, implementation of the cohesive zone model requires interface elements to be placed along the path of fracture.
This requires anticipating the crack path, and could potentially lead to mesh dependency \cite{zhou2004dynamic}.

Phase field fracture (PFF) methods are another class of techniques that replace a sharp crack with a diffused crack field $\eta(\bm{x},t)$ with an associated length scale $\eta_\varepsilon$ \cite{kuhn2010continuum,bourdin2008variational,del2013variational}.
The crack field is set to $1$ inside the crack, and zero outside. 
PFF methods introduce an energy contribution corresponding to the crack field which governs crack field evolution.
Additionally, a regularization term is introduced that circumvents the mesh dependency of the solution.
The PFF solution has been mathematically shown to converge to the analytical linear elastic fracture mechanics (LEFM) solution in the limit $\eta_\varepsilon\rightarrow 0$ \cite{bourdin2008variational}.

Central to all of these methods is the computation of a stress field around a crack tip, which always produces highly localized strain fields.
In each of these cases, the mesh must be highly refined near the crack tip but can be coarse in the far field.
In some cases this is accomplished by completely re-meshing at various intervals or by pre-meshing in anticipation of the crack's path \cite{lee2016pressure,schrefler2006adaptive,khoei2008modeling}.
Here, we apply the finite difference BSAMR linear solver to the problem of linear elastic fracture mechanics for use with PFF.

The energy functional $\mathcal{F}$ is written as
\begin{equation}
  \mathcal{F} = \int_\Omega g(\eta) W(\nabla\bm{u}) d\Omega + \int_\Omega \mathcal{G}_c \left[ \frac{w(\eta)}{\eta_\varepsilon} + \eta_\varepsilon |\nabla\eta|^2 \right]\,d\Omega
  \label{eq:PFFCrackEnergy}
\end{equation}
where $\mathcal{G}_c$ is the crack fracture energy, $\bm{u}$ is the displacement, $W(\nabla \bm{u}) = \mathbb{C}_{ijkl}u_{i,j}u_{k,l}$ is the elastic strain energy and $\eta_\varepsilon$ is the length scale associated with diffused crack width.
Additionally, the functions $g(\eta)$ and $w(\eta)$ are differentiable functions chosen such that $g(\eta=1) =1 $ and $w(\eta=1) = 0$.
In this case, $g(\eta) = \eta^2$ and $w(\eta) = 1 - g(\eta)$. 
The last term in equation (\ref{eq:PFFCrackEnergy}) is for crack regularization that eliminates mesh dependency of the solution.

\begin{figure}[t]
    \centering
    \includegraphics[width=0.5\linewidth]{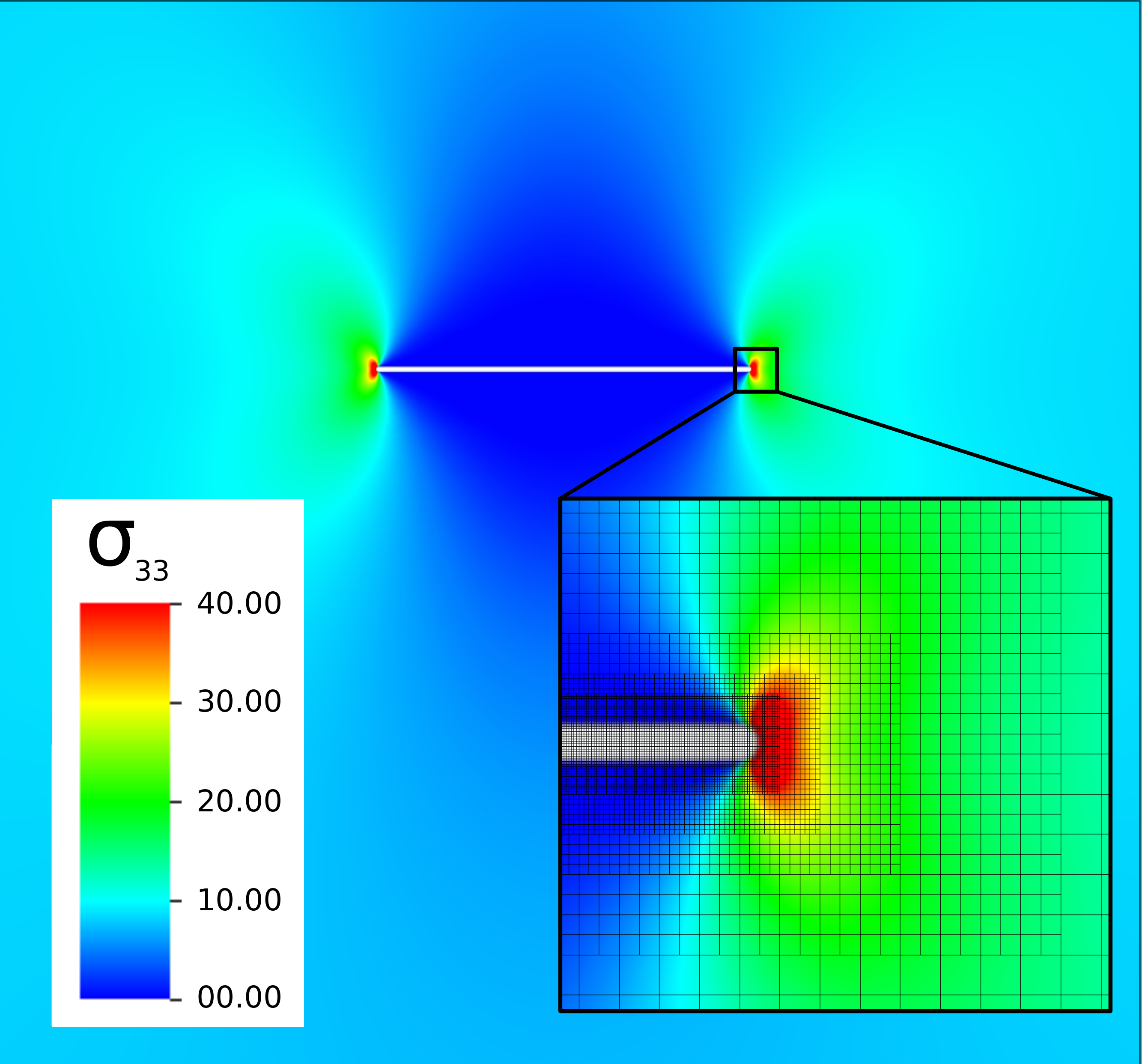}
    \caption{(Fracture) Far field stress ($\sigma_{22}$) field (main) and detail of stress field near crack tip (inset). }
    \label{fig:fracture_sigma22_field}
\end{figure}

The equilibrium equations for $\bm{u}$ and $\eta$ are obtained using the variational derivative of $\mathcal{F}$ with $\bm{u}$ and $\eta$ respectively.
It can be shown that the stress equilibrium equation is
\begin{equation}
  \text{div}\left(g(\eta) \frac{\partial W}{\partial \left( \nabla\bm{u} \right)}\right) = 0
\end{equation}
A two dimensional rectangular domain $(-4,-4)\times(4,4)$ is chosen with a notch of length $a$, thickness $t$ and semi-circular ends of radius $t/2$, centered at $(0,0)$.
The boundary is mollified with length scale $\varepsilon$ using the error function approach described in equation (\ref{eq:ErrorFunctionMollification}). 
To solve the problem, a base mesh of $512\times 512$ points was chosen with $5$ levels of refinement. 
The criterion
\begin{align}
    |\nabla c(\bm{x})|\,|\Delta\bm{x}| > 0.001
\end{align}
was used to tag cells for refinement. 
The elastic modulus was capped to $1\%$ inside the crack to maintain numerical stability.
Computations were performed with MPI using 16 cores.
The stress distribution near the crack tip is shown in Figure \ref{fig:fracture_sigma22_field}.

\begin{figure}[t]
    \centering
    \begin{subfigure}{0.5\linewidth}
        \includegraphics[width=\linewidth]{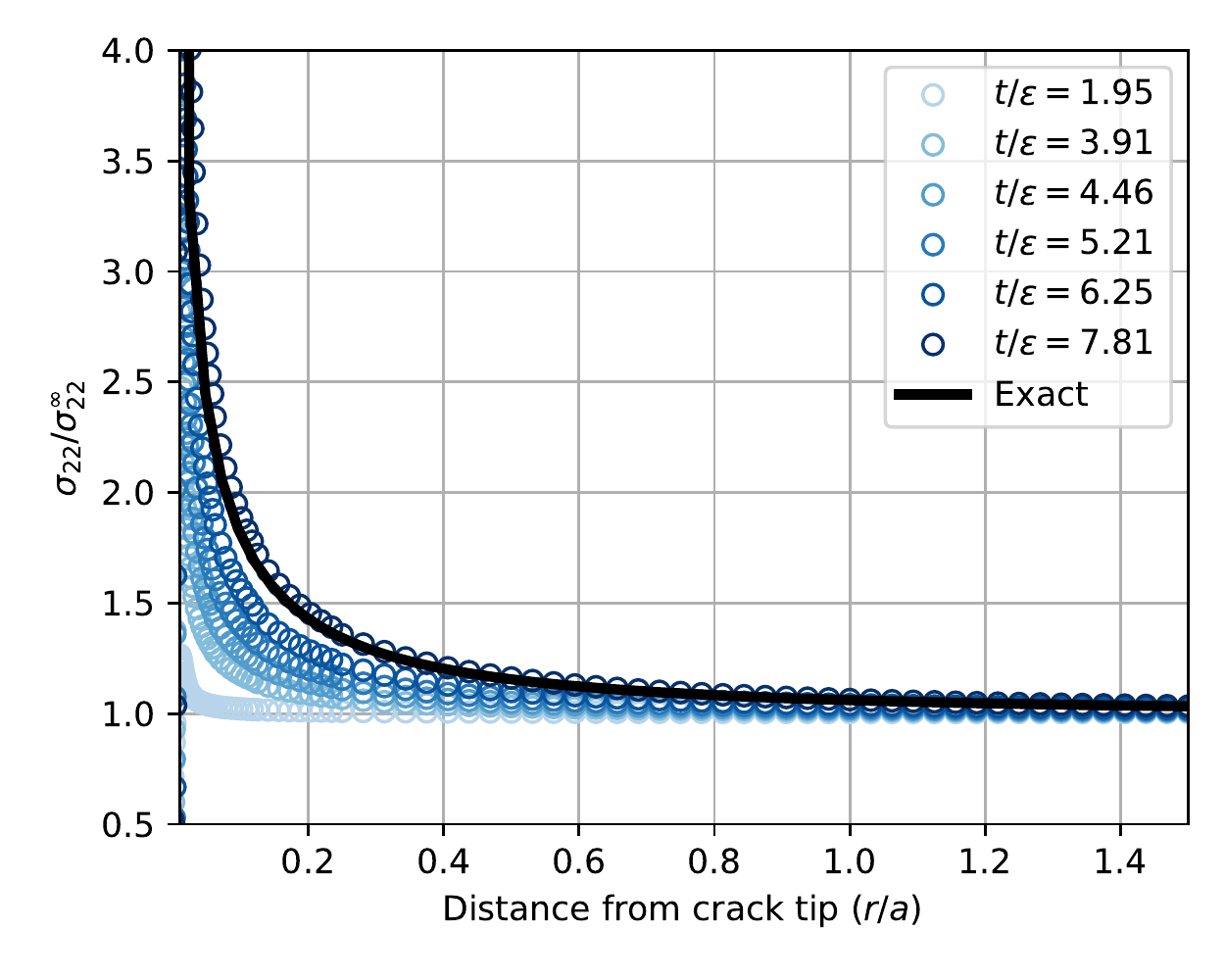}
        \caption{Stress contour}
    \end{subfigure}%
    \begin{subfigure}{0.5\linewidth}
        \includegraphics[width=\linewidth]{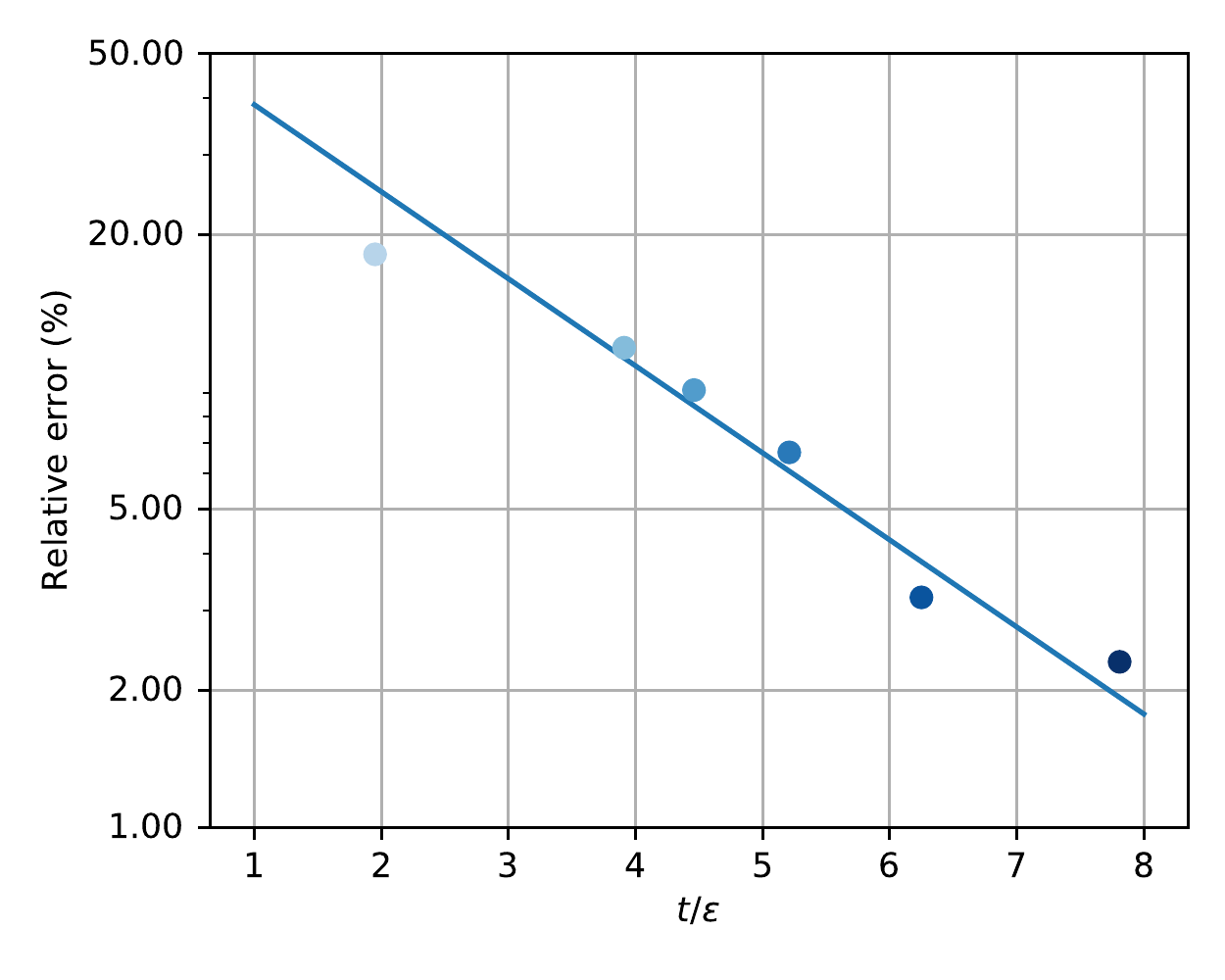}
        \caption{Relative error}
    \end{subfigure}
    
    \caption{(Fracture) Variation of crack stress $\sigma_{22}$ with $r$ at $\theta=0$, for fixed crack length to thickness ratio and different mollification length scales. The analytical solution is compared to the data.}
    \label{fig:fracture_sigma22}
\end{figure}

To validate the BSAMR approach, stress $\sigma_{22}$ is plotted along $y=0$ line for $a/t = 64$ and different mollification length scales. 
The results are compared to analytical LEFM solutions provided in \cite{tada2000analysis,lacazette1990application} in Figure \ref{fig:fracture_sigma22}.
It can be observed that as mollification length scale reduces, the predicted crack stresses approach the analytical solution (Figure~\ref{fig:fracture_sigma22}).
Figure \ref{fig:fracture_sigma22}b shows the decrease in relative error as $\varepsilon$ decreses.

\subsection{Phase field microstructure}

\begin{figure}
  \centering
  \includegraphics[width=0.5\linewidth]{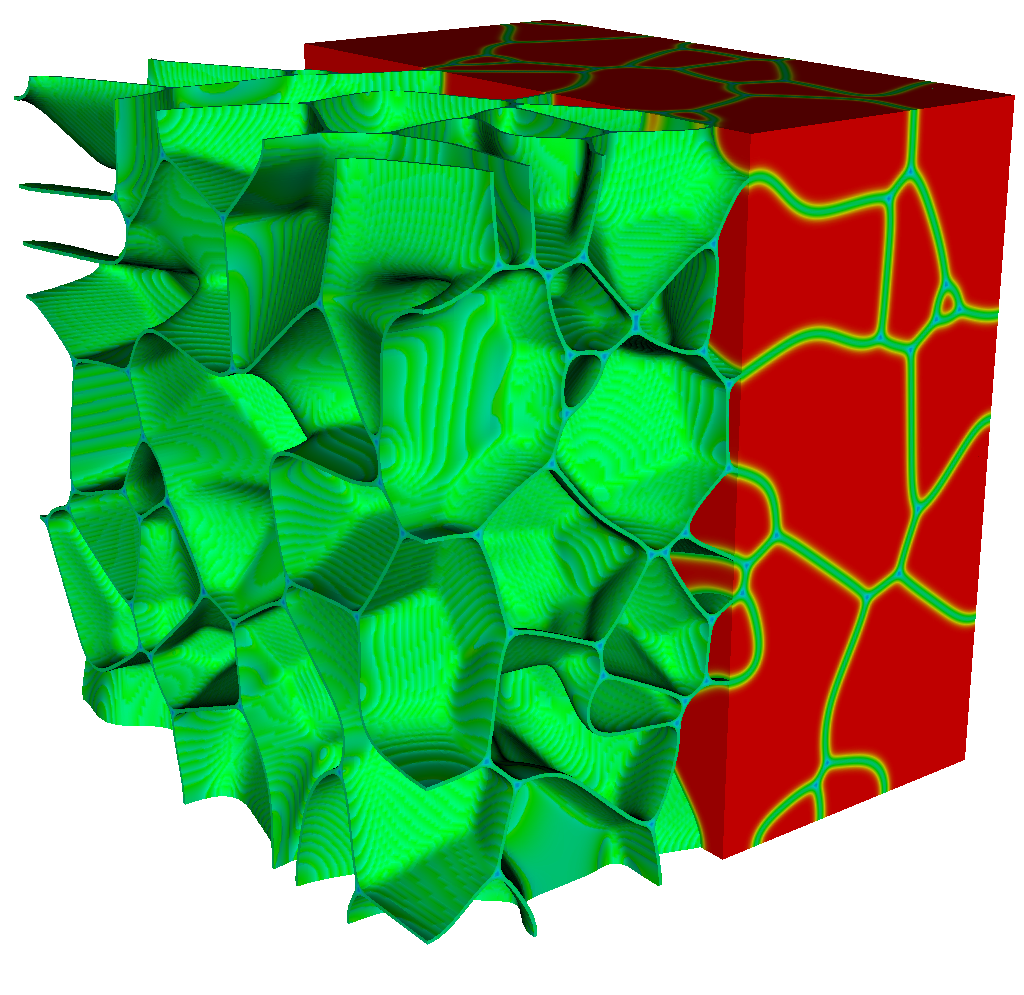}
  \caption{(Microstructure) Plot of $\sum_i\eta_i\eta_i$ highlighting boundary regions - isosurfaces (left) and full field (right)}
  \label{fig:microstructure_eta}
\end{figure}

A primary motivational application of interest for the present work is RVE mesoscale simulations of microstructural response.
The multiphase field (MPF) method is used to evolve boundaries.
Though the MPF method itself is a focus of this work, we present a brief overview here.
Let a polycrystal with $N$ grains be represented by $N$ order functions called order parameters $\{\eta_1,\ldots,\eta_N\}\subset C(\Omega,[0,1])$.
The order parameters act as indicator functions for the regions occupied for each grain; grain $n$ occupies the region in which $\eta_n=1$.
The boundaries are diffuse, such that there is a smooth transition region between grain with a characteristic length $\ell$ (Figure~\ref{fig:microstructure_eta}). 
The order parameters are evolved using the differential equations

\begin{align}\label{eq:allen_cahn}
  \frac{\partial\eta^n}{\partial t} = -L\,\frac{\delta F}{\delta t}
\end{align}

where $F(\eta_1\ldots\eta_N,\nabla\eta_1\ldots\eta_N)$ is a free energy functional.
For more details on the method see \cite{moelans2008introduction,moelans2011quantitative,moelans2008phase,ribot2019new}.

An example of an implementation using the base form of $F$ is discussed in \cite{ribot2019new}, although we do not consider the higher-order regularzation here.
To account for mechanically-driven grain boundary motion, the functional $F$ also contains an elastic energy term:

\begin{align}
  \frac{1}{2}\bm{\varepsilon}:\mathbb{C}(\eta_1,\ldots,\eta_N):\bm{\varepsilon},
\end{align}
where $\bm{\varepsilon}$ are the solutions to the linear elasticity problem and the elastic modulus tensor $\mathbb{C}$ is given by a simple mixture rule
\begin{align}
  \mathbb{C}(\eta_1,\ldots,\eta_N) = 
    \frac{1}{\sum_{i=1}^N\eta_i} \sum_{i=1}^N\,\mathbb{C}_i\,\eta_i.
\end{align}
Other mixture rules can be used (c.f. \cite{moelans2011quantitative}), but a linear rule is sufficient for purposes of model demonstration.

The unrotated fourth order elastic modulus tensor is expressed in Voigt notation as 

\begin{align}
  \mathbb{C} =
  \begin{bmatrix}
    C_{11} & C_{12} & C_{12} & 0     & 0     & 0      \\
    C_{12} & C_{11} & C_{12} & 0     & 0     & 0      \\
    C_{12} & C_{12} & C_{11} & 0     & 0     & 0      \\
    0     & 0     & 0     & C_{44} & 0     & 0      \\
    0     & 0     & 0     & 0     & C_{44} & 0      \\
    0     & 0     & 0     & 0     & 0     & C_{44}  
  \end{bmatrix}
\end{align}
which solves $\bm{\sigma}=\mathbb{C}\bm{\varepsilon}$ when 

\begin{gather}
  \bm{\sigma}=\begin{bmatrix}\sigma_{11}&\sigma_{22}&\sigma_{33}&\sigma_{23}&\sigma_{31}&\sigma_{12}\end{bmatrix} \text{, and } \\
  \bm{\varepsilon}=\begin{bmatrix}\varepsilon_{11}&\varepsilon_{22}&\varepsilon_{33}&2\varepsilon_{23}&2\varepsilon_{31}&2\varepsilon_{12}\end{bmatrix}.
\end{gather}
Each crystal is given a random orientation $\bm{R}\in SO(3)$, and the rotated elastic modulus tensor $\hat{\mathbb{C}}$ is 

\begin{align}
  \hat{\mathbb{C}}_{pqst} = R_{pi}R_{qj}\mathbb{C}_{ijkl}R^T_{ks}R^T_{lt}.
\end{align}
To retain all crystallographic information, it is necessary to retain, at minimum, 6 constants: the moduli $C_{11},C_{12},C_{44}$, and the 3 degrees of freedom associated with the rotation.
However, storing only these constants would require an extensive number of trigonometric calculations every time the stress is calculated, which would be computationally intensive.
Instead, all 21 components (the symmetric part of $\mathbb{C}$) are stored at each point.

\begin{figure}[h]\centering
  \begin{subfigure}{0.33\linewidth}
    \caption{$\sigma_{xx}$}
    \includegraphics[width=\linewidth]{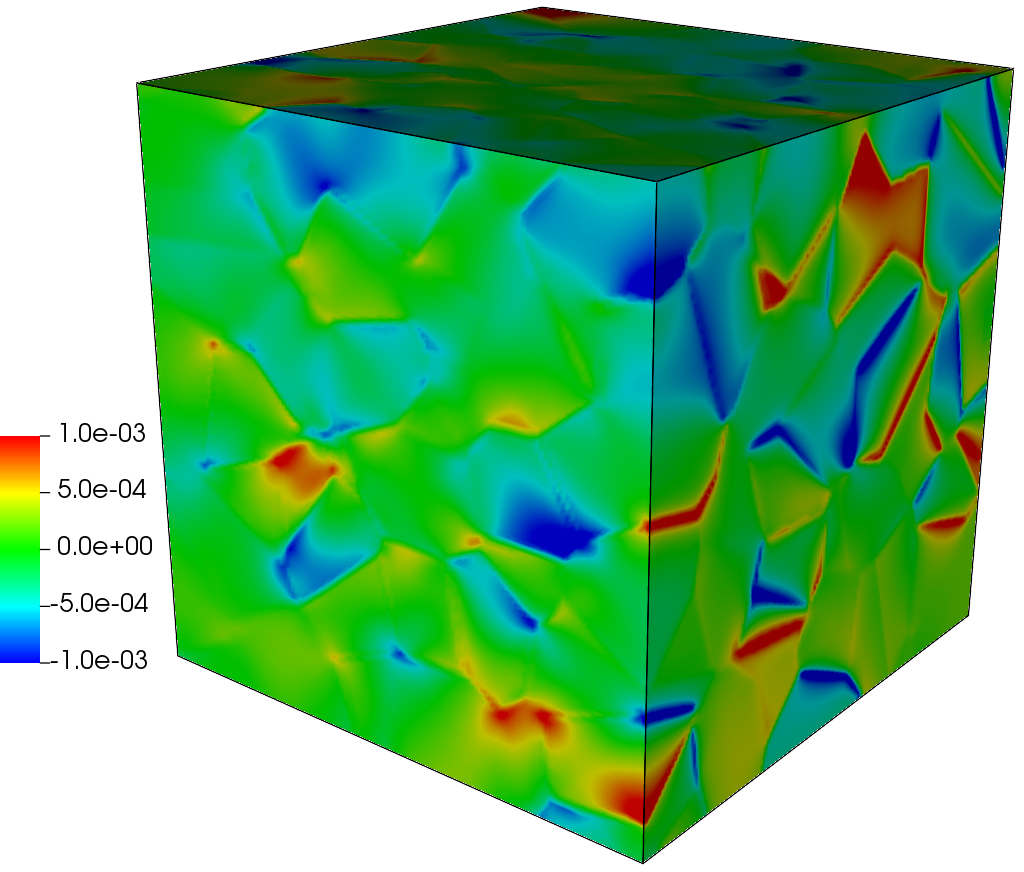}
  \end{subfigure}%
  \begin{subfigure}{0.33\linewidth}
    \caption{$\sigma_{yy}$}
    \includegraphics[width=\linewidth]{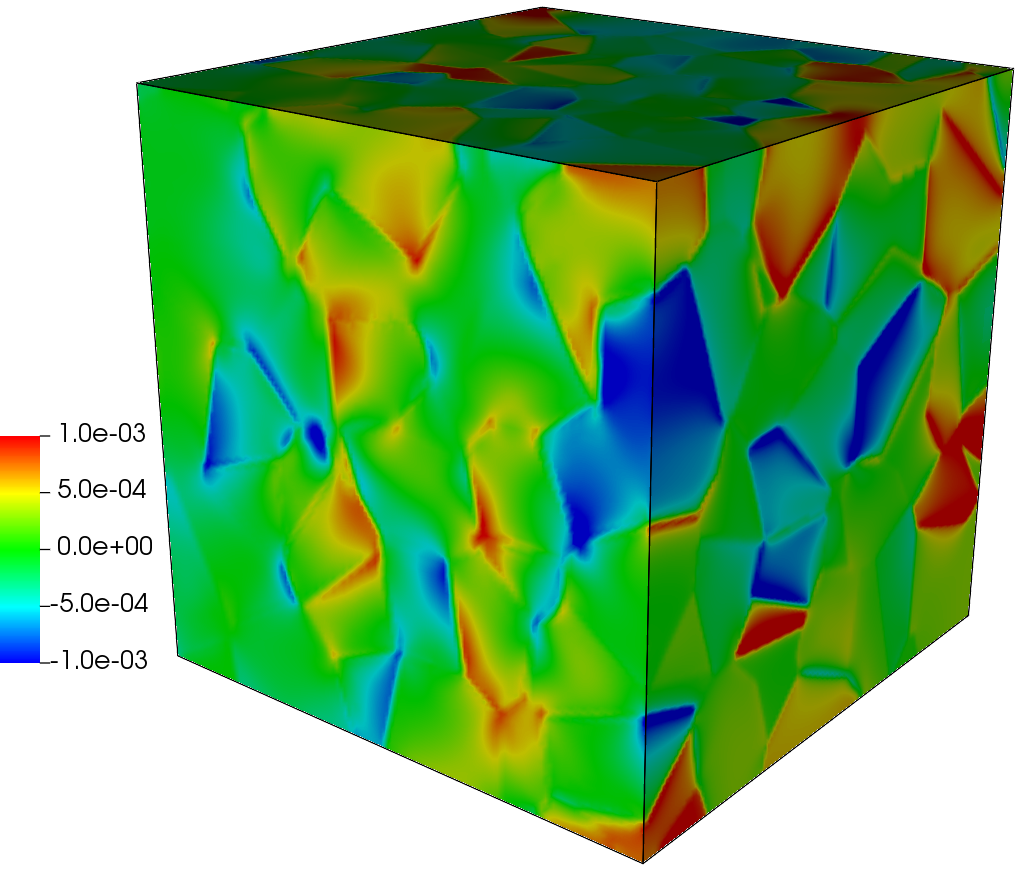}
  \end{subfigure}%
  \begin{subfigure}{0.33\linewidth}
    \caption{$\sigma_{zz}$}
    \includegraphics[width=\linewidth]{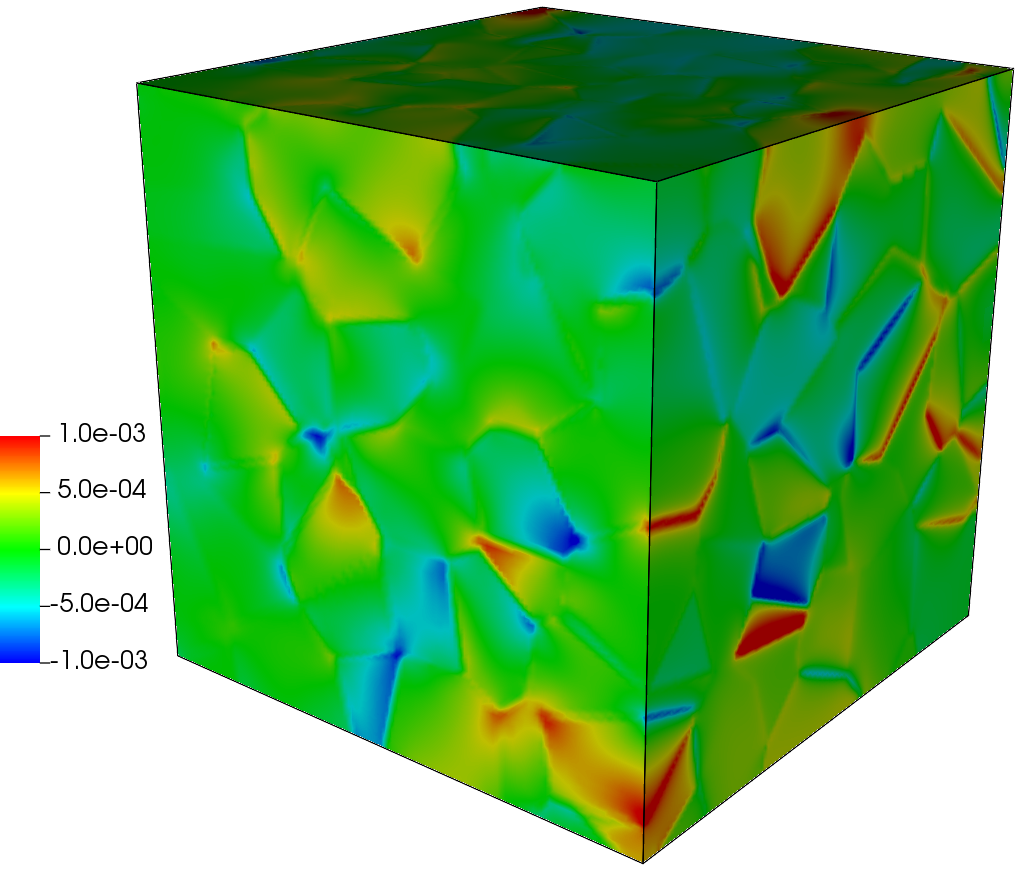}
  \end{subfigure}
  \begin{subfigure}{0.33\linewidth}
    \caption{$\sigma_{yz}$}
    \includegraphics[width=\linewidth]{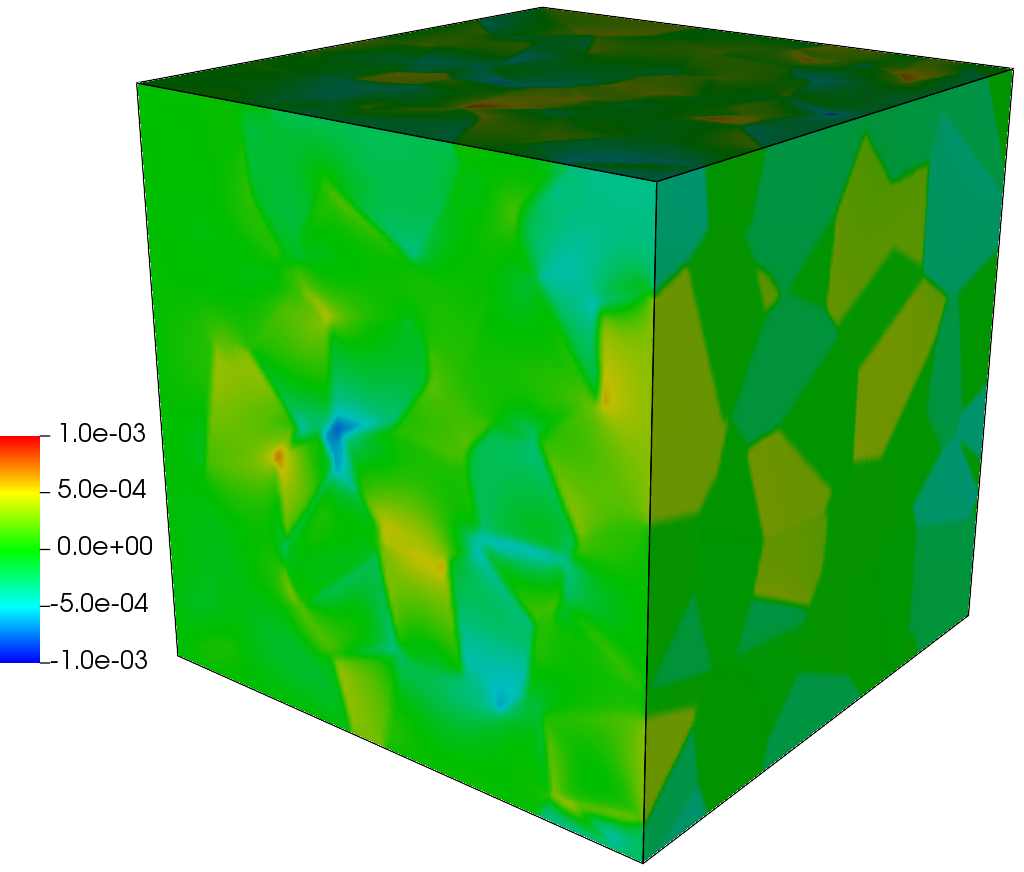}
  \end{subfigure}%
  \begin{subfigure}{0.33\linewidth}
    \caption{$\sigma_{zx}$}
    \includegraphics[width=\linewidth]{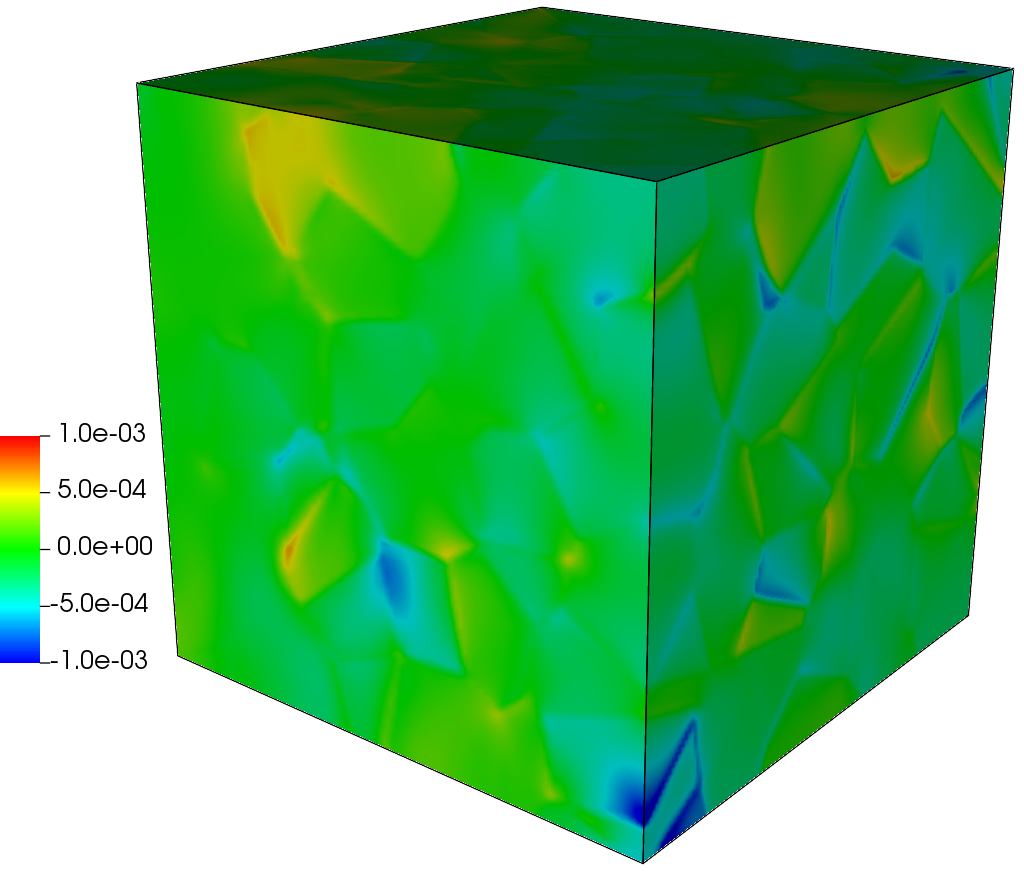}
  \end{subfigure}%
  \begin{subfigure}{0.33\linewidth}
    \caption{$\sigma_{xy}$}
    \includegraphics[width=\linewidth]{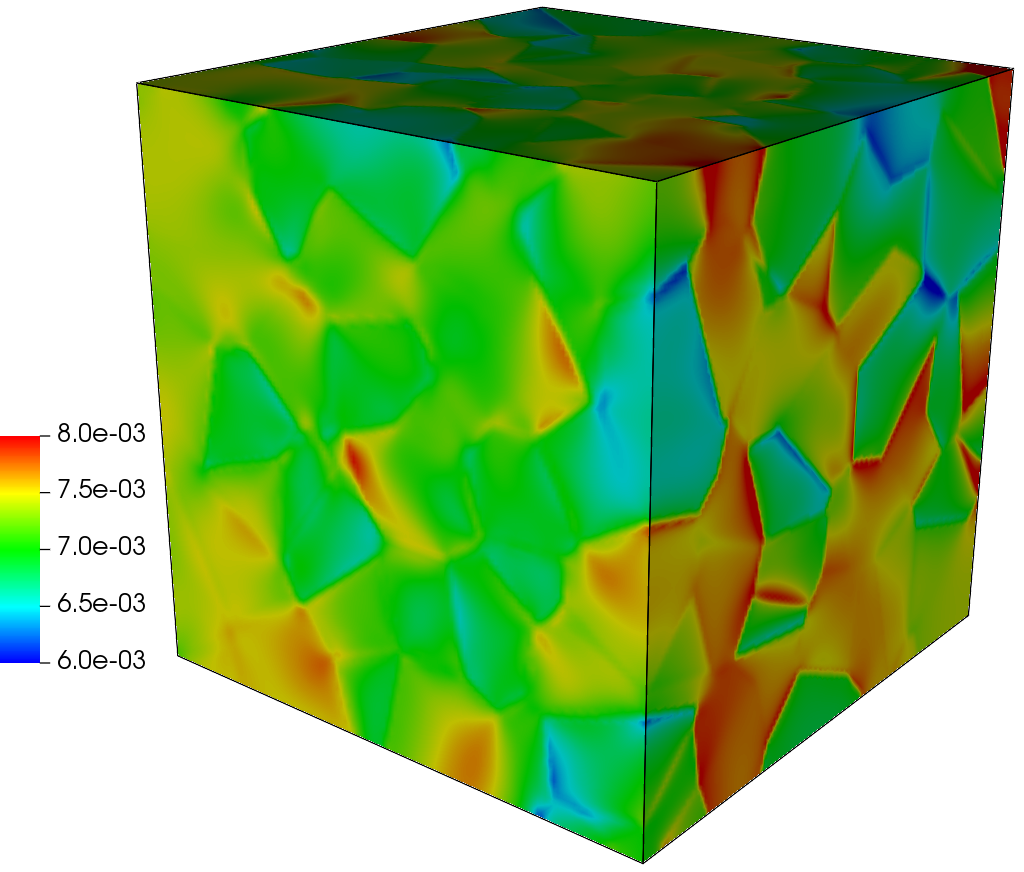}
  \end{subfigure}
  \caption{(Microstructure) Plot of components of $\sigma$}
  \label{fig:microstructure_sigma_comp}
\end{figure}

\begin{figure}[h]\centering
    \includegraphics[width=0.5\linewidth]{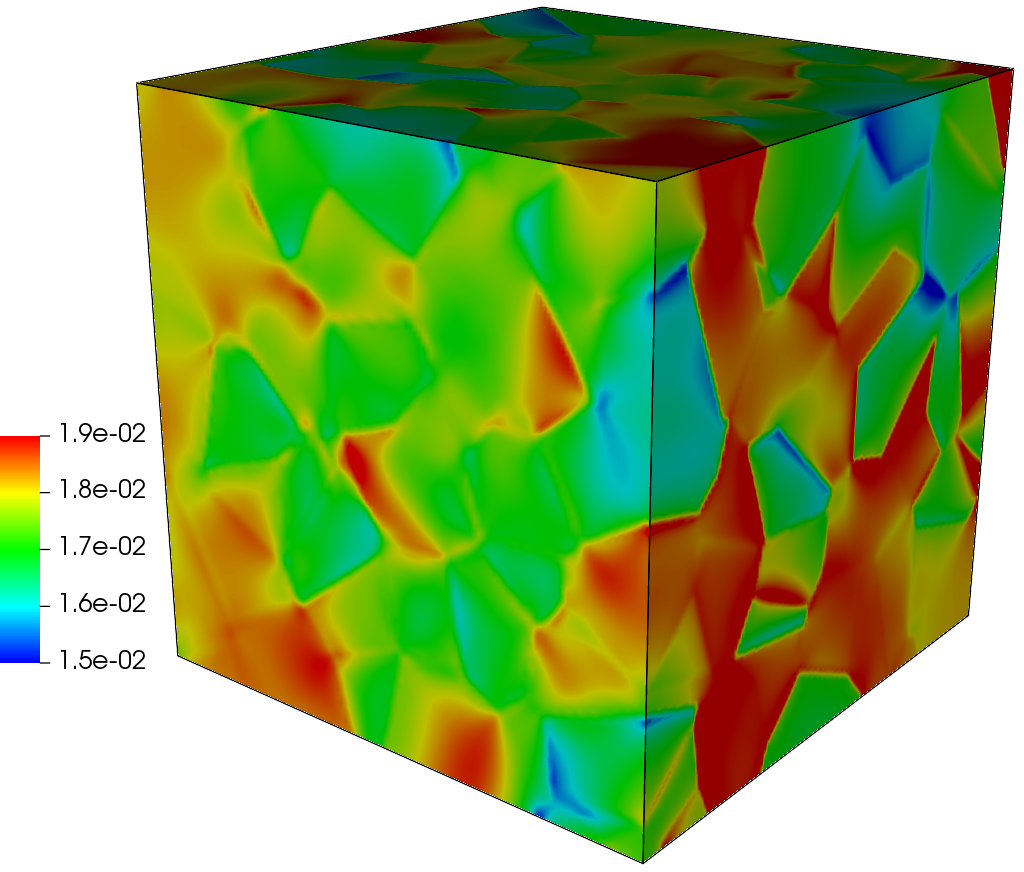}
  \caption{(Microstructure) Von Mises stress}
  \label{fig:microstructure_sigma_vm}
\end{figure}

To set up the simulation, a cubic RVE is used and initialized using a Voronoi tessellation.
For all simulations, a total of 10 order parameters are used; however, the order parameters are re-used, and the Voronoi tessellation initializes up to 400 grains.
Each order parameter is initialized with a randomly selected orientation.
The evolution equation (\ref{eq:allen_cahn}) is integrated for 10 timesteps, which allows the grain boundaries to sufficiently diffuse.
This is necessary in order to avoid mesh-size dependent effects resulting from aliasing at the grain boundaries.

For the phase field evolution, the following parameters are used.
Grain boundary mobility is $M=1.0$, and the constants for the chemical potential are $\mu=10.0,\gamma=1.0$.
The diffuse grain boundary width is $\ell_{GB}=0.025$ to ensure sufficient resolution.
The grain boundary energy is $\sigma_0=0.075$.
The three cubic elastic constants are $C_{11}=1.68, C_{12}=1.21, C_{44}=0.75$.
The spatial dimension for the simulation domain are $\Omega = [0,10]^3 \,/\,[0,5]^3$.
A base grid (refinement level 0) is $64^3$ (large) and $32^3$ (small).
The criterion
\begin{equation}
  \bigvee_{i=1}^N |\nabla\eta_i(\bm{x})|\,|\Delta\bm{x}| > 0.1
\end{equation}
was used to tag cells for refinement.
For the large simulations, 400 grains were used, for the small, 10 grains.
For the large simulations, 3 AMR levels were used, for the small, 4 AMR levels were used.
For the multigrid solver, 6 levels were used for the large grid and 5 for the small.

Once the boundaries have evolved and diffused, a mechanical loading boundary condition is applied.
The top is subjected to a shear displacement of $0.1$, and the displacement field is relaxed using the AMR multi-level multi-grid solver.\par

The solution is a complex stress field resulting from the crystallographic anisotropy and heterogeneity resulting from the polycrystalline structure.
The six stress components (normal stresses $\sigma_{xx},\sigma_{yy},\sigma_{zz}$, shear stresses $\sigma_{yz},\sigma_{zx},\sigma_{xy}$) are plotted in Figure~\ref{fig:microstructure_sigma_comp}.
A full analysis of the the accuracy of these results is outside the scope of this work, due to the inherent difficulty in matching 3D stress fields for Voronoi tessellations to 2D microstructure. 
Therefore, the results here are presented as a proof of the applicability of this method to polycrystalline material science applications, rather than an example of validation.
(For comparable polycrystalline elasticity simulations using other methods, the reader is referred to \cite{zeghadi2007ensemble,brenner2009elastic,vidyasagar2017predicting}.)
The von-Mises stress is defined as

\begin{align}
  \sigma_{VM}=\sqrt{\frac{3}{2}S_{ij}S_{ij}},
\end{align}
where $S_{ij}$ is the stress deviator tensor.
$\sigma_{VM}$ is plotted in Figure~\ref{fig:microstructure_sigma_vm}, and indicates (unsurprisingly) that the maximum von Mises stress occurs at the grain boundaries.

The solver generally converges in no more than 36 iterations, and this number remained constant regardless of problem size or scaling.
Similar tests have been run for isotropic materials (in which the isotropic constants, rather than the crystallographic orientation, varies between grains) and the solver consistently converges in about half the number of iterations.
This difference is most likely due to the inefficacy of the Jacobi preconditioner on cubic materials compared to isotropic; more optimal preconditioners can be used for further performance enhancement.
The microstructure calculations were used as a benchmark for performance, which is discussed in detail in the following section.

\section{Performance}\label{sec:performance}

The Alamo implementation of the finite difference reflux-free AMR method was tested for BSAMR efficiency and massive parallelism.
In all of the tests presented here, only the time spent in the linear solver is reported.
The time spent initializing the Voronoi microstructure, and the time required to evolve the phase field order parameters, are not included in the performance analysis.

\subsection{AMR performance}

2D tests are run on the phase field microstructure and Eshelby inclusion problems to determine the effect of AMR on the solver efficiency. 
All tests are run on a laptop using no more than 6 MPI processes. 

\begin{figure}[t]
  \centering
  \begin{minipage}{0.5\linewidth}
  \includegraphics[width=0.32\linewidth]{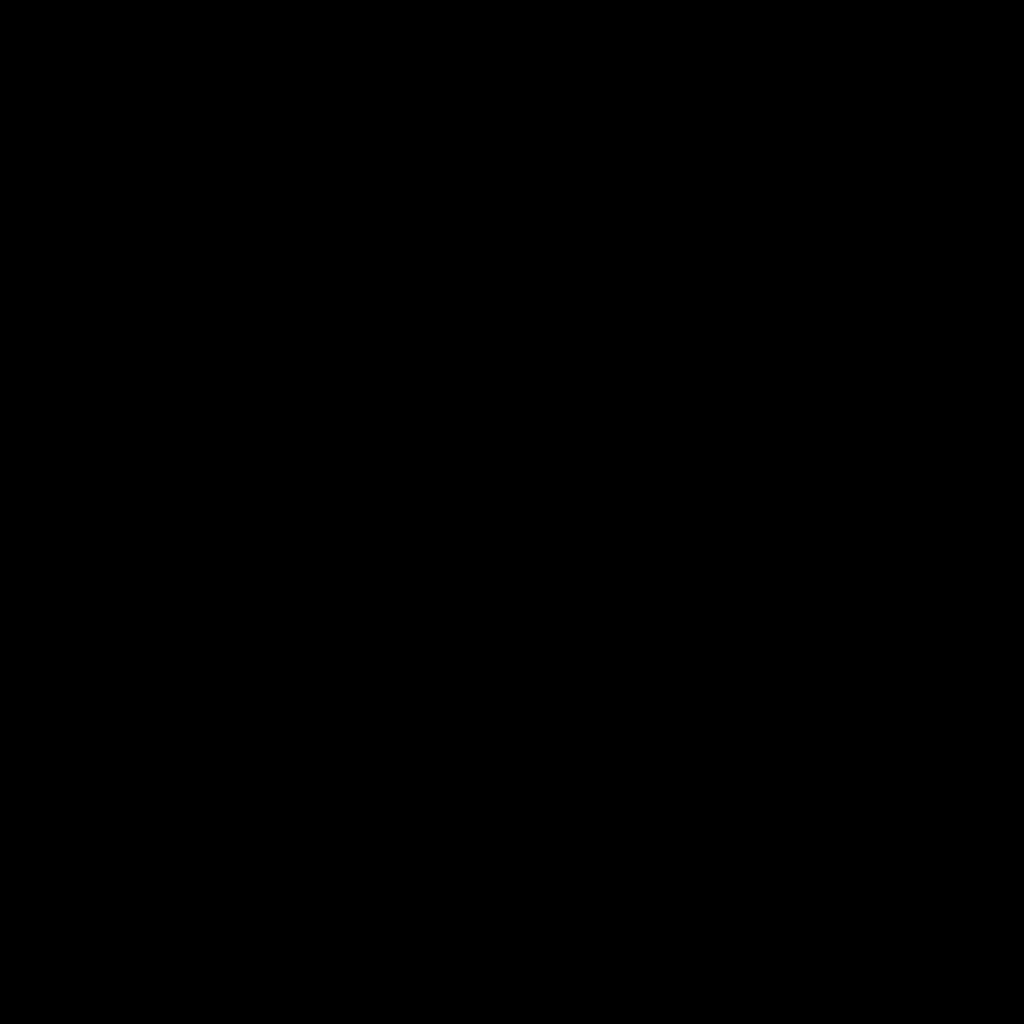}%
  \includegraphics[width=0.32\linewidth]{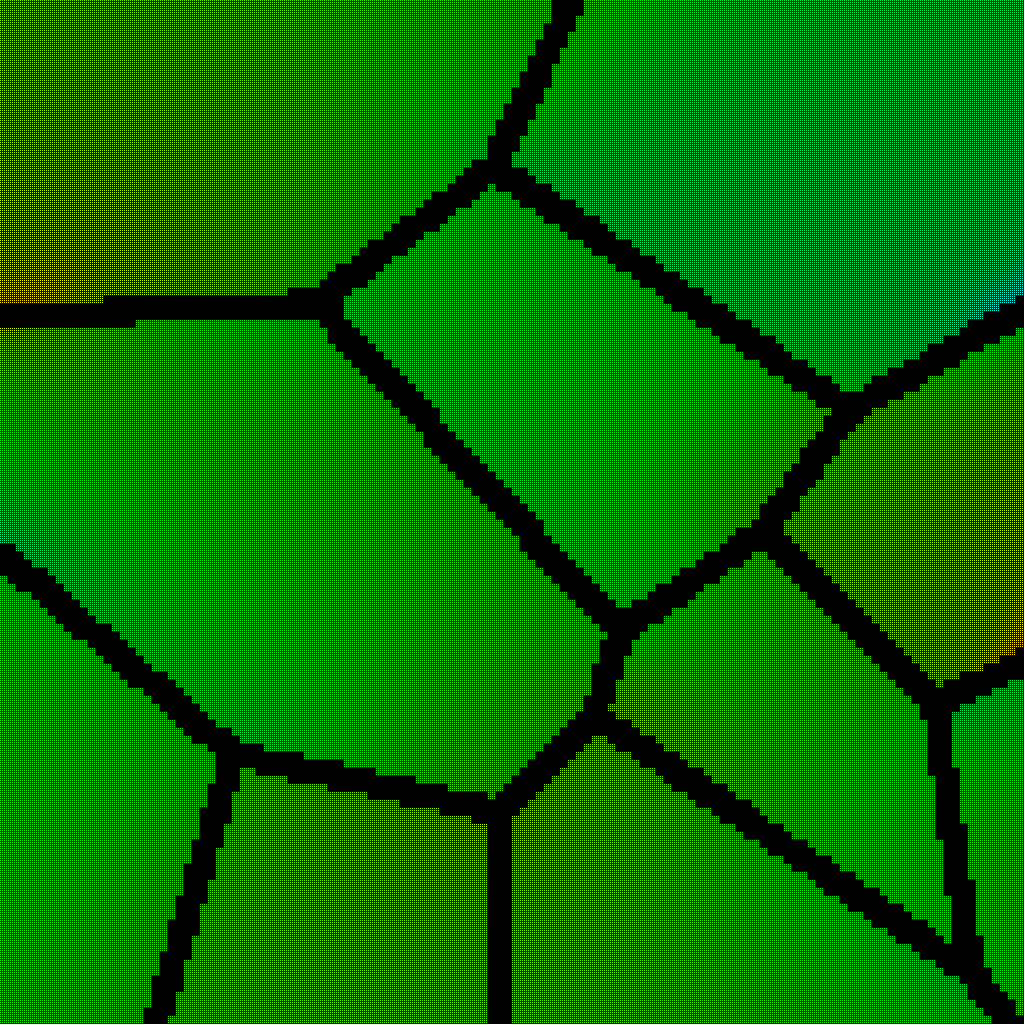}%
  \includegraphics[width=0.32\linewidth]{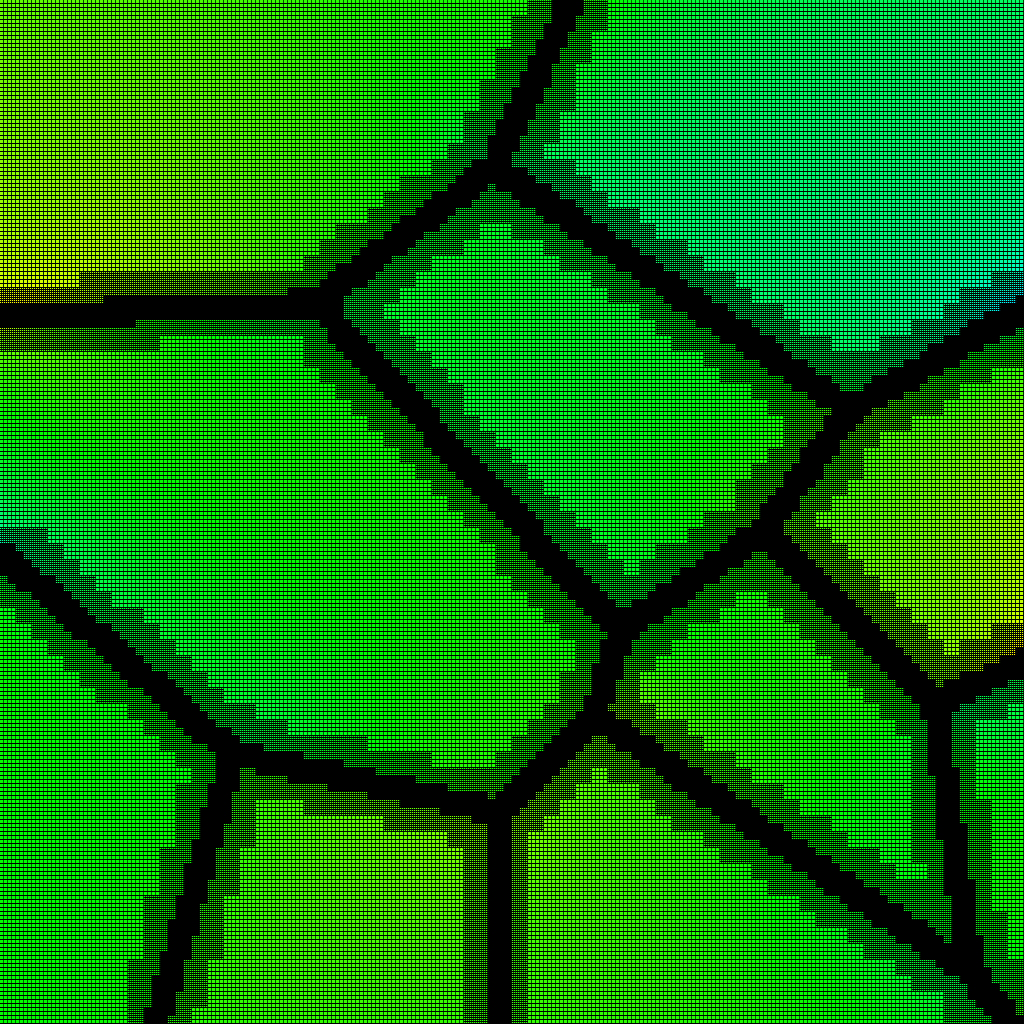}
  \includegraphics[width=0.32\linewidth]{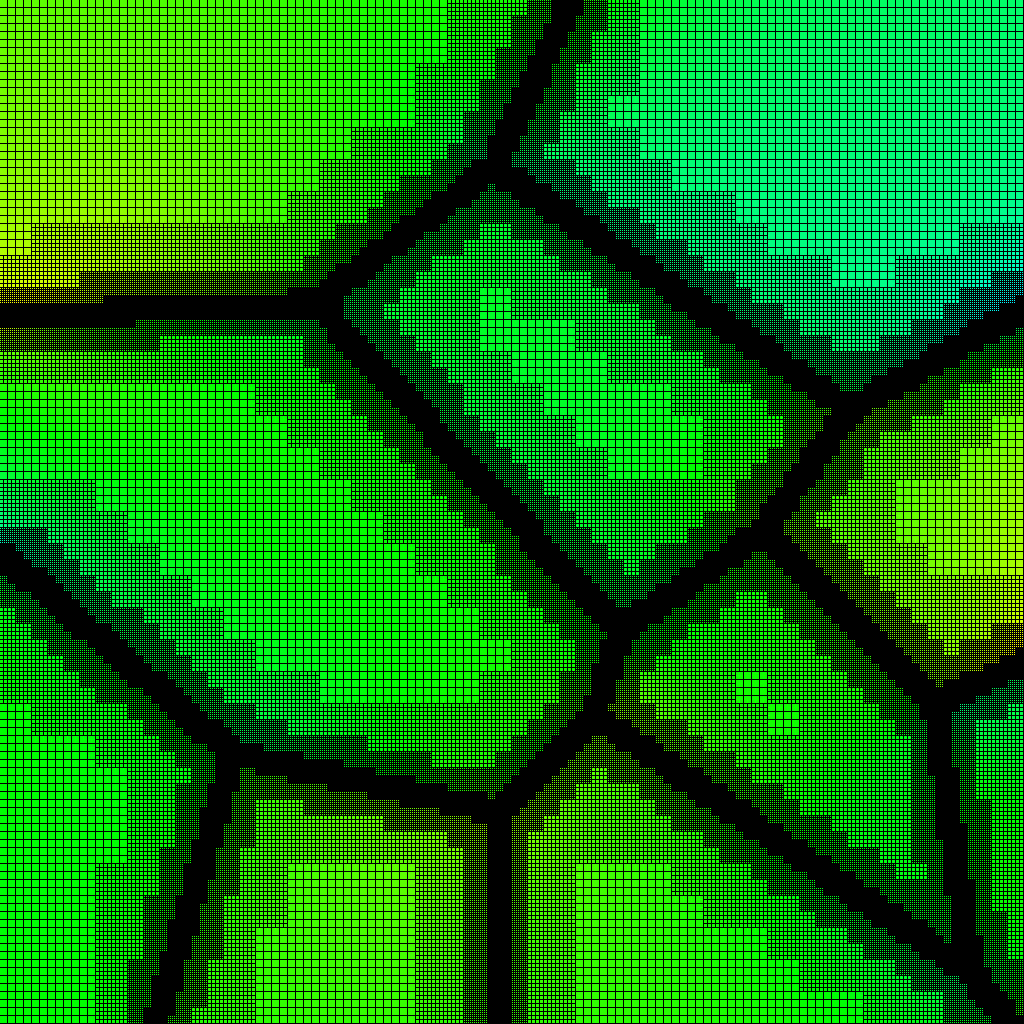}%
  \includegraphics[width=0.32\linewidth]{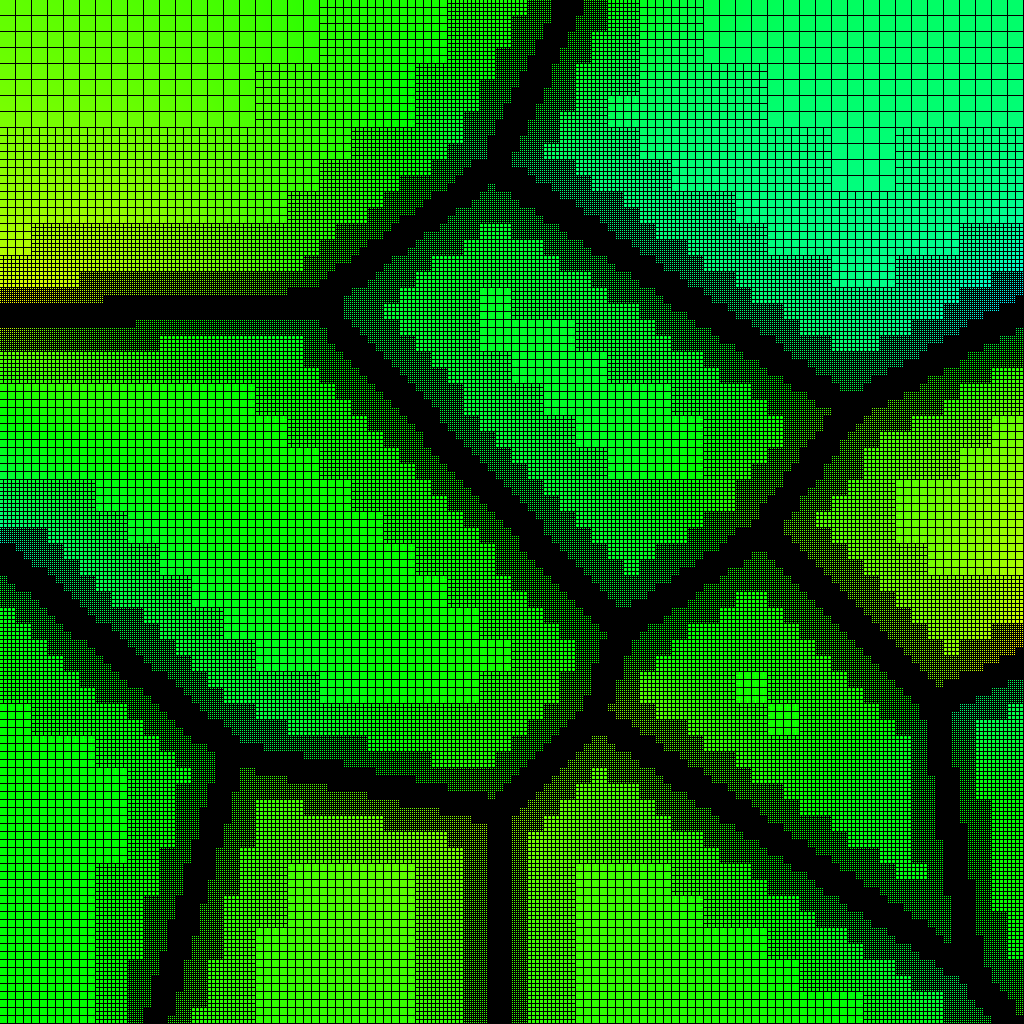}%
  \includegraphics[width=0.32\linewidth]{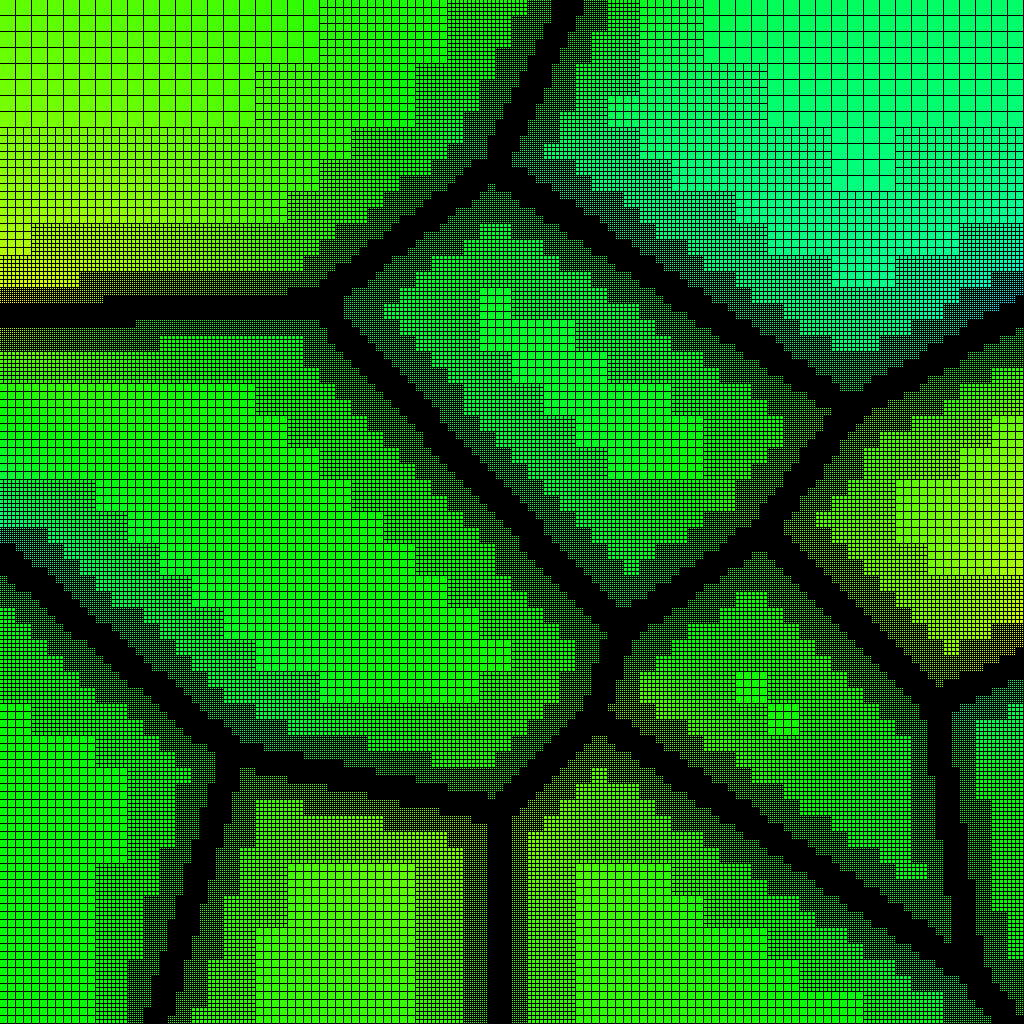}
  \end{minipage}
  \caption{
    Meshes with varying levels of refinement, beginning with no refinement (upper left).
  }
  \label{fig:successive-amr}
\end{figure}

For each case, a simulation is performed at maximum resolution for the entire domain, with no AMR.
As additional AMR levels are added, the maximum resolution remains the same, so that the final results are identical in accuracy.
Figure~\ref{fig:successive-amr} illustrates this process for phase field microstructure, applying up to 5 levels of AMR.
It can be seen that there is initially a dramatic decrease in the number of nodes for the first 2-3 AMR applications; however, there is no difference between the 5-level case and the 4-level case as it is not possible to re-grid without violating the AMR conditions.
We refer to the point at which additional AMR levels are not beneficial as ``AMR saturation.''

For the phase field microstructure case, AMR saturation is reached at 2-3 refinement levels.
For the Eshelby inclusion case, it is reached at 4-5. 
This is nothing other than a function of the geometry of the problem, the optimization of which is outside the scope of this paper. 

\begin{figure}[h]
  \centering
  \includegraphics[width=0.5\linewidth]{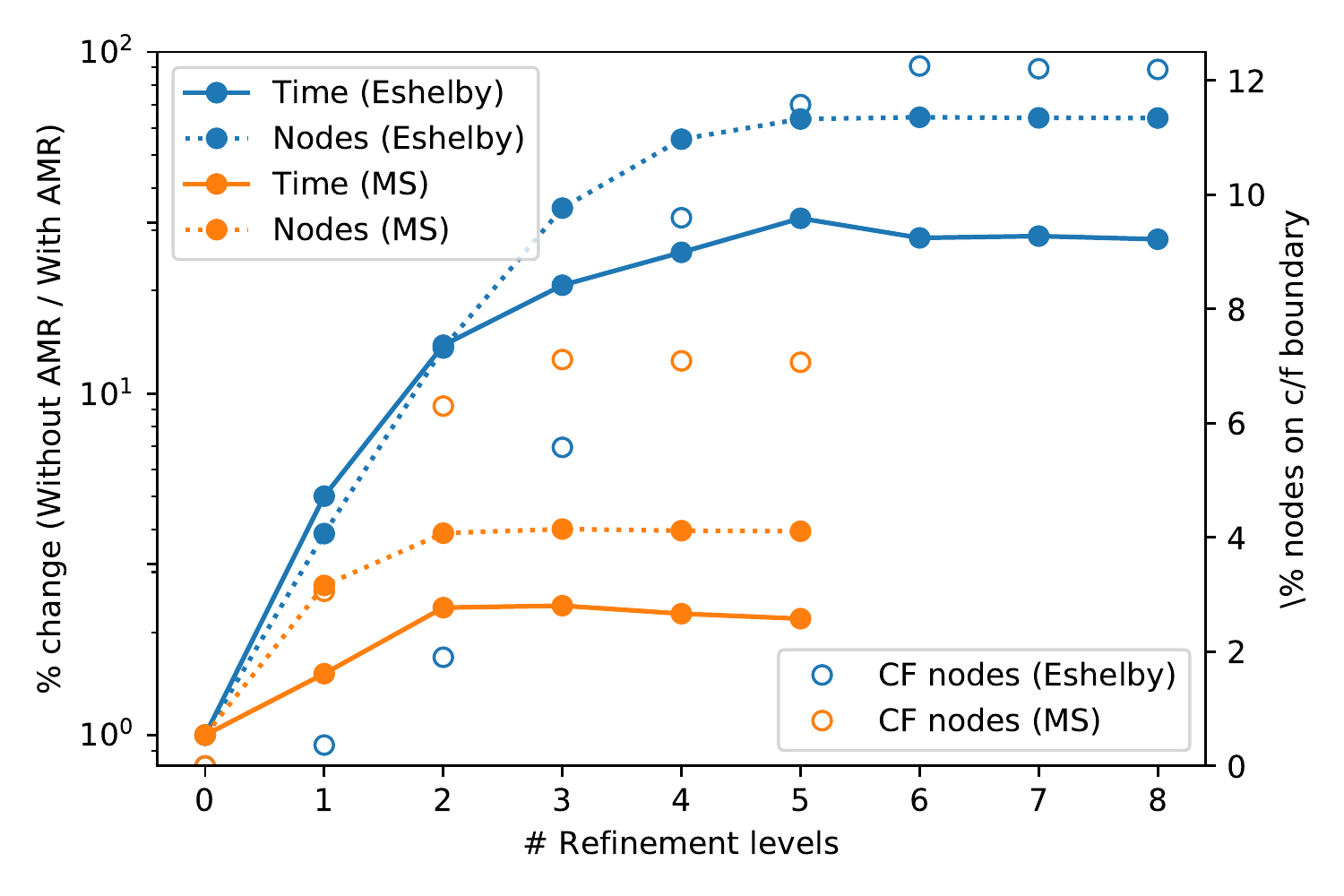}
  \caption{
    Performance data for Eshelby (blue) and Microstructure (gold) cases.
    Solid lines represent the no-AMR solver time divided by the AMR solver time; dashed lines represent the number of nodes on the fully-refined grid divided by the number of nodes on the AMR grid. 
    Unfilled markers indicate the percentage of the computational domain on the coarse/fine boundary.
  }
  \label{fig:performance}
\end{figure}

Solver data is collected for each of these simulations at each level of refinement (Figure~\ref{fig:performance})
For the Eshelby case, the processing time scaled nearly perfectly with the number of nodes, for a maximum speedup of approximately 6,600\%. 
This can be attributed to the extreme well-suitedness of the Eshelby problem to AMR: the region of interest is extremely small compared to the domain size, and is confined to a nearly spherical region. 

The phase field microstructure results similarly indicate that the solver performance scales similarly with the node count, although the scaling is not nearly as close as that for the Eshelby case, only achieving a speedup of about 230\%.
This is to be expected, as the refined region for the microstructure case is non-localized, resulting in a large amount of coarse/fine interface.

\subsection{Parallel scaling}

Large and small-scale strong scaling tests were performed using the STAMPEDE2 supercomputer at the Texas Advanced Computing Center.
STAMPEDE2 has 1,736 compute nodes with Intel Xeon Platinum 8160 (``Skylake'') nodes; each has two sockets with 24 cores per socket.

\begin{figure}
  \centering\includegraphics[width=0.5\linewidth]{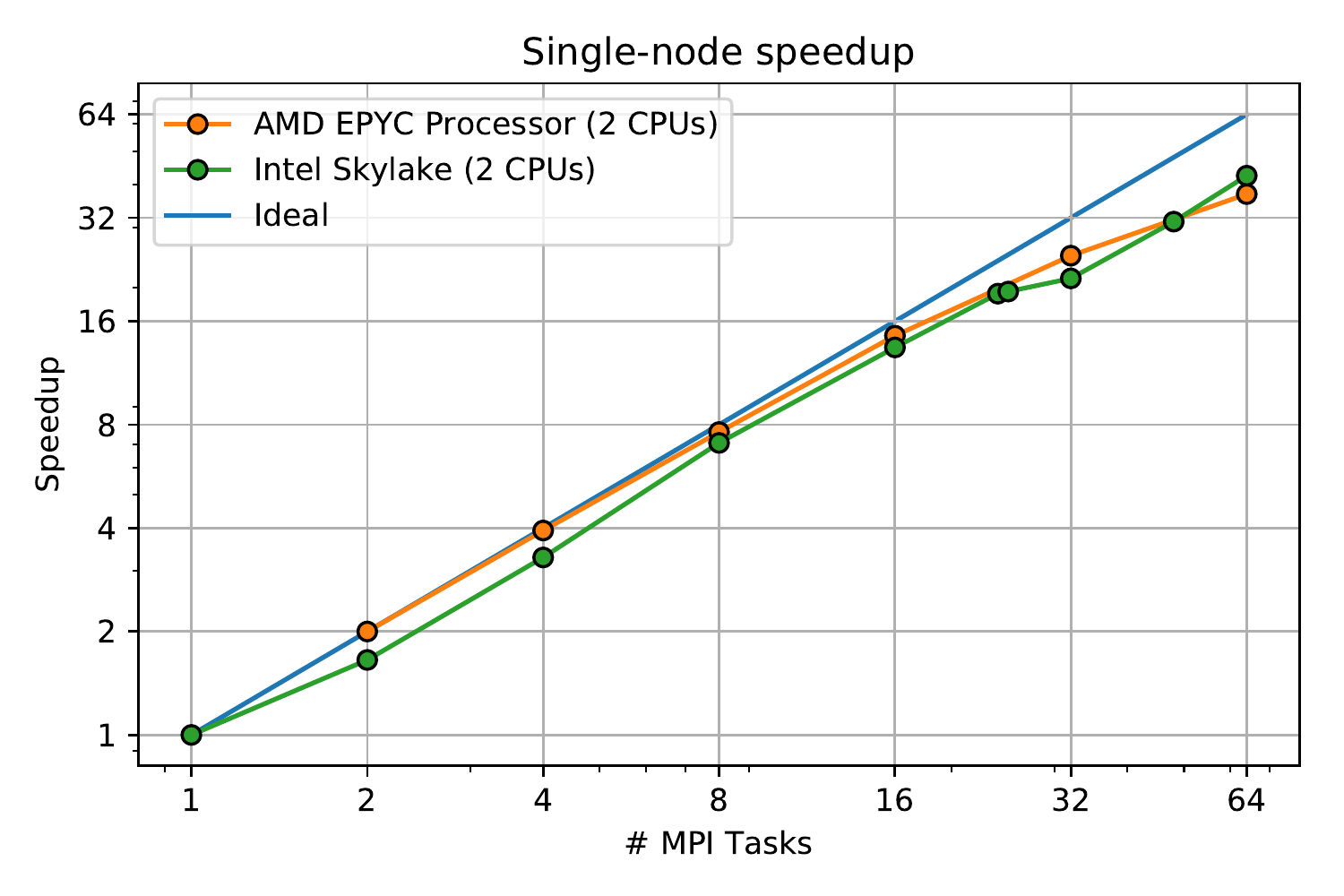}
  \caption{Speedup measurements with 10 grains, 7.6 million grid points on Intel Skylake processors (green) and AMD EPYC processors.}
  \label{fig:microstructure_n000010_speedup}
\end{figure}

For small-scale, single-node scaling tests, the small microstructure parameters were used.
Speedup results are shown in Figure~\ref{fig:microstructure_n000010_speedup}.
For 1-16 processors, nearly perfect speedup is observed.
This is unsurprising as the number of MPI processes is less than the number of cores per socket, and so the entire computation is on a single socket.
Once the number of processors exceeds 24, there is a dip in performance below ideal speedup due to the overhead resulting from inter-node communication.
Nevertheless, speedup continues to increase at nearly the same rate.
The minimum computation time for the solver with 48 MPI tasks was approximately 8.3 minutes with 7,640,576 total grid points distributed among 4 AMR levels.
On the finest AMR level, 5,588,480 nodes were concentrated on 33\% of the simulation domain.
(Note: time for initialization, phase field evolution, and plot file I/O is not included here.) 

Small-scale single-node scaling tests are repeated on a small workstation using AMD EPYC CPUs (Figure~\ref{fig:microstructure_n000010_speedup}).
The simulation involves 10 grains, with four AMR levels subject to an elastic loading. 
The workstation used has 2x AMD EPYC 7282 processors (16 core/32 thread), with 64 GB RAM.
It is observed that the performance drops when simulation is run for all 64 threads.
This performance decrease is attributed to machine overhead effects that come into picture when 100\% of the machine is being used.

\begin{figure}
  \centering\includegraphics[width=0.5\linewidth]{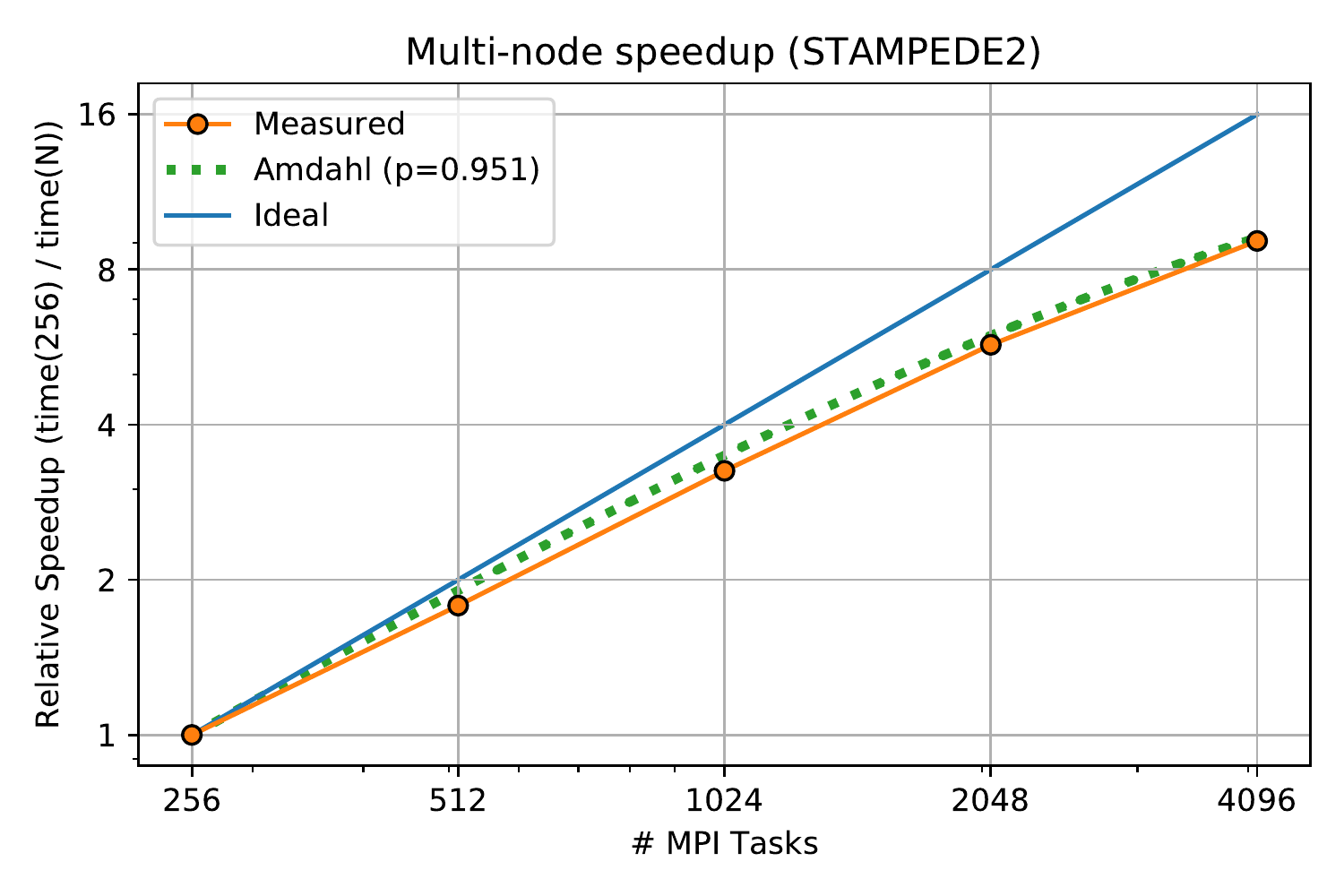}
  \caption{
    Relative speedup measurements with 400 grains, 102.1 million grid points, on the STAMPEDE2 supercomputer.
    (Relative speedup = time with N processors / time with 256 processors)
  }
  \label{fig:microstructure_N000400_speedup}
\end{figure}

For large-scale, multi-node scaling tests, the large microstructure parameters are used.
The only difference was the use of 4 AMR levels instead of 3.
Relative speedup results (i.e. speedup based on the smallest number of MPI tasks) are plotted in Figure~\ref{fig:microstructure_N000400_speedup}.
Amdahl's speedup curve is included as well, with an experimentally measured value of $p=0.951$.

\section{Conclusions and Acknowledgments}\label{sec:conclusions}

In this work we have developed and tested a finite difference elastic solver on block-structured AMR grids with geometric multigrid.
While FEA is a mainstay of computational solid mechanics, BSAMR is an application in which the traditional advantages of FEM are no longer applicable.
For computational problems that use BSAMR, the proposed method has shown to be a feasible alternative.
Verification was provided by comparison to the standard Eshelby inclusion case, as well as the fracture mechanics case.
It was shown that the method is capable of resolving a stress field around an approximated void with a diffuse boundary.
Finally, the method was applied to large microstructure evolution problems, and shown to produce qualitatively reasonable results.
More importantly, the method was shown to exhibit good scaling, making it a feasible method for computational micromechanics.

There are some limitations to the proposed method.
First, it is limited to meshes composed only of regular grids.
Problems with complex domain geometry would require special treatment (for instance, using embedded boundaries), and are less amenable to the proposed approach. 
(Such problems would be more well-suited to FEM.)
Second, it requires all material discontinuities to be diffused in order to avoid a singularity resulting from the second derivative.
Therefore, the primary domain of application of this method is those cases in which these requirements are already satisfied.

BR acknowledges support from Lawrence Berkeley National Laboratory (LBL) subcontract \#7473053, as well as some support from the UCCS Committee for Research and Creative Works.
VA acknowledges Hopper computing cluster at Auburn University.
WQ and AA acknowledge support from the U.S. Department of Energy, Office of Science, Office of Advanced Scientific Computing Research, Applied Mathematics Program under Award No. DE-SC0008271 and under Contract No. DE-AC02-05CH11231.
The HPC parts of this work were made possible through the use of the STAMPEDE2 supercomputer at the Texas Advanced Computing Center (TACC), through an allocation provided by the Extreme Science and Engineering Discovery Environment (XSEDE) \cite{towns2014xsede}, allocation \# TG-PHY130007.
XSEDE is supported by National Science Foundation grant number ACI-1548562.

\bibliographystyle{ieeetr}
\bibliography{library}

\begin{thebibliography}{10}

\bibitem{meyers1994dynamic}
M.~A. Meyers, {\em Dynamic behavior of materials}.
\newblock John wiley \& sons, 1994.

\bibitem{keshavarz2013multi}
S.~Keshavarz and S.~Ghosh, ``Multi-scale crystal plasticity finite element
  model approach to modeling nickel-based superalloys,'' {\em Acta Materialia},
  vol.~61, no.~17, pp.~6549--6561, 2013.

\bibitem{keshavarz2015hierarchical}
S.~Keshavarz and S.~Ghosh, ``Hierarchical crystal plasticity fe model for
  nickel-based superalloys: sub-grain microstructures to polycrystalline
  aggregates,'' {\em International Journal of Solids and Structures}, vol.~55,
  pp.~17--31, 2015.

\bibitem{jafari2017constitutive}
M.~Jafari, M.~Jamshidian, S.~Ziaei-Rad, D.~Raabe, and F.~Roters, ``Constitutive
  modeling of strain induced grain boundary migration via coupling crystal
  plasticity and phase-field methods,'' {\em International Journal of
  Plasticity}, vol.~99, pp.~19--42, 2017.

\bibitem{vedantam2006efficient}
S.~Vedantam and B.~Patnaik, ``Efficient numerical algorithm for multiphase
  field simulations,'' {\em Physical Review E}, vol.~73, no.~1, p.~016703,
  2006.

\bibitem{schmitz2010phase}
G.~Schmitz, B.~B{\"o}ttger, J.~Eiken, M.~Apel, A.~Viardin, A.~Carr{\'e}, and
  G.~Laschet, ``Phase-field based simulation of microstructure evolution in
  technical alloy grades,'' {\em International Journal of Advances in
  Engineering Sciences and Applied Mathematics}, vol.~2, no.~4, pp.~126--139,
  2010.

\bibitem{plewa2005adaptive}
T.~Plewa, T.~Linde, and V.~G. Weirs, ``Adaptive mesh refinement-theory and
  applications,'' {\em Lecture notes in computational science and engineering},
  vol.~41, pp.~3--5, 2005.

\bibitem{dubey2014survey}
A.~Dubey, A.~Almgren, J.~Bell, M.~Berzins, S.~Brandt, G.~Bryan, P.~Colella,
  D.~Graves, M.~Lijewski, F.~L{\"o}ffler, {\em et~al.}, ``A survey of high
  level frameworks in block-structured adaptive mesh refinement packages,''
  {\em Journal of Parallel and Distributed Computing}, vol.~74, no.~12,
  pp.~3217--3227, 2014.

\bibitem{neiva2019scalable}
E.~Neiva, S.~Badia, A.~F. Mart{\'\i}n, and M.~Chiumenti, ``A scalable parallel
  finite element framework for growing geometries. application to metal
  additive manufacturing,'' {\em International Journal for Numerical Methods in
  Engineering}, vol.~119, no.~11, pp.~1098--1125, 2019.

\bibitem{morales2013numerical}
J.~Morales, J.~Moreno, and F.~Alhama, ``Numerical solution of 2d elastostatic
  problems formulated by potential functions,'' {\em Applied Mathematical
  Modelling}, vol.~37, no.~9, pp.~6339--6353, 2013.

\bibitem{harangus2014finite}
K.~Harangus and A.~Kakucs, ``Finite-difference solution using displacement
  potential function for plane stresses and displacements,'' {\em Procedia
  Technology}, vol.~12, pp.~394--400, 2014.

\bibitem{chern1986front}
I.-L. Chern, J.~Glimm, O.~McBryan, B.~Plohr, and S.~Yaniv, ``Front tracking for
  gas dynamics,'' {\em Journal of Computational Physics}, vol.~62, no.~1,
  pp.~83--110, 1986.

\bibitem{terashima2009front}
H.~Terashima and G.~Tryggvason, ``A front-tracking/ghost-fluid method for fluid
  interfaces in compressible flows,'' {\em Journal of Computational Physics},
  vol.~228, no.~11, pp.~4012--4037, 2009.

\bibitem{agrawal2018impact}
V.~Agrawal and K.~Bhattacharya, ``Impact induced depolarization of
  ferroelectric materials,'' {\em Journal of the Mechanics and Physics of
  Solids}, vol.~115, pp.~142--166, 2018.

\bibitem{sethian2001evolution}
J.~A. Sethian, ``Evolution, implementation, and application of level set and
  fast marching methods for advancing fronts,'' {\em Journal of computational
  physics}, vol.~169, no.~2, pp.~503--555, 2001.

\bibitem{gibou2018review}
F.~Gibou, R.~Fedkiw, and S.~Osher, ``A review of level-set methods and some
  recent applications,'' {\em Journal of Computational Physics}, vol.~353,
  pp.~82--109, 2018.

\bibitem{beyer1992computational}
R.~P. Beyer~Jr, ``A computational model of the cochlea using the immersed
  boundary method,'' {\em Journal of computational physics}, vol.~98, no.~1,
  pp.~145--162, 1992.

\bibitem{li1994immersed}
Z.~Li, {\em The immersed interface method: a numerical approach for partial
  differential equations with interfaces}.
\newblock PhD thesis, 1994.

\bibitem{roma1999adaptive}
A.~M. Roma, C.~S. Peskin, and M.~J. Berger, ``An adaptive version of the
  immersed boundary method,'' {\em Journal of computational physics}, vol.~153,
  no.~2, pp.~509--534, 1999.

\bibitem{peskin2002immersed}
C.~S. Peskin, ``The immersed boundary method,'' {\em Acta numerica}, vol.~11,
  pp.~479--517, 2002.

\bibitem{zhang2019amrex}
W.~Zhang, A.~Almgren, V.~Beckner, J.~Bell, J.~Blaschke, C.~Chan, M.~Day,
  B.~Friesen, K.~Gott, D.~Graves, {\em et~al.}, ``Amrex: a framework for
  block-structured adaptive mesh refinement,'' {\em Journal of Open Source
  Software}, 2019.

\bibitem{briggs2000multigrid}
W.~L. Briggs, S.~F. McCormick, {\em et~al.}, {\em A multigrid tutorial},
  vol.~72.
\newblock Siam, 2000.

\bibitem{almgren1998conservative}
A.~S. Almgren, J.~B. Bell, P.~Colella, L.~H. Howell, and M.~L. Welcome, ``A
  conservative adaptive projection method for the variable density
  incompressible navier--stokes equations,'' {\em Journal of computational
  Physics}, vol.~142, no.~1, pp.~1--46, 1998.

\bibitem{stolarski1983shear}
H.~Stolarski and T.~Belytschko, ``Shear and membrane locking in curved c0
  elements,'' {\em Computer methods in applied mechanics and engineering},
  vol.~41, no.~3, pp.~279--296, 1983.

\bibitem{li2010comparison}
S.~Li, ``Comparison of refinement criteria for structured adaptive mesh
  refinement,'' {\em Journal of computational and applied mathematics},
  vol.~233, no.~12, pp.~3139--3147, 2010.

\bibitem{sverdrup2018highly}
K.~Sverdrup, N.~Nikiforakis, and A.~Almgren, ``Highly parallelisable
  simulations of time-dependent viscoplastic fluid flow simulations with
  structured adaptive mesh refinement,'' {\em arXiv preprint arXiv:1803.00417},
  2018.

\bibitem{ream2019numerical}
J.~Ream, M.~Henry~de Frahan, M.~Martin, S.~Yellapantula, and R.~W. Grout,
  ``Numerical simulations of the supercritical carbon dioxide round turbulent
  jet,'' tech. rep., National Renewable Energy Lab.(NREL), Golden, CO (United
  States), 2019.

\bibitem{vay2018warp}
J.-L. Vay, A.~Almgren, J.~Bell, L.~Ge, D.~Grote, M.~Hogan, O.~Kononenko,
  R.~Lehe, A.~Myers, C.~Ng, {\em et~al.}, ``Warp-x: A new exascale computing
  platform for beam--plasma simulations,'' {\em Nuclear Instruments and Methods
  in Physics Research Section A: Accelerators, Spectrometers, Detectors and
  Associated Equipment}, vol.~909, pp.~476--479, 2018.

\bibitem{nonaka2019amrex}
A.~J. Nonaka, ``The amrex block structured adaptive mesh refinement library:
  Astrophysical applications,'' in {\em American Astronomical Society Meeting
  Abstracts\# 233}, vol.~233, 2019.

\bibitem{fullmer2019benchmarking}
W.~D. Fullmer, A.~S. Almgren, M.~Rosso, J.~Blaschke, and J.~Musser,
  ``Benchmarking of a preliminary mfix-exa code,'' {\em arXiv preprint
  arXiv:1909.02067}, 2019.

\bibitem{eshelby1957determination}
J.~D. Eshelby, ``The determination of the elastic field of an ellipsoidal
  inclusion, and related problems,'' {\em Proceedings of the Royal Society of
  London. Series A. Mathematical and Physical Sciences}, vol.~241, no.~1226,
  pp.~376--396, 1957.

\bibitem{eshelby1959elastic}
J.~D. Eshelby, ``The elastic field outside an ellipsoidal inclusion,'' {\em
  Proceedings of the Royal Society of London. Series A. Mathematical and
  Physical Sciences}, vol.~252, no.~1271, pp.~561--569, 1959.

\bibitem{jin2016explicit}
X.~Jin, D.~Lyu, X.~Zhang, Q.~Zhou, Q.~Wang, and L.~M. Keer, ``Explicit
  analytical solutions for a complete set of the eshelby tensors of an
  ellipsoidal inclusion,'' {\em Journal of Applied Mechanics}, vol.~83, no.~12,
  p.~121010, 2016.

\bibitem{torries2018overview}
B.~Torries, A.~Imandoust, S.~Beretta, S.~Shao, and N.~Shamsaei, ``Overview on
  microstructure-and defect-sensitive fatigue modeling of additively
  manufactured materials,'' {\em Jom}, vol.~70, no.~9, pp.~1853--1862, 2018.

\bibitem{park2011cohesive}
K.~Park and G.~H. Paulino, ``Cohesive zone models: a critical review of
  traction-separation relationships across fracture surfaces,'' {\em Applied
  Mechanics Reviews}, vol.~64, no.~6, p.~060802, 2011.

\bibitem{jin2006comparison}
Z.-H. Jin and C.~Sun, ``A comparison of cohesive zone modeling and classical
  fracture mechanics based on near tip stress field,'' {\em International
  journal of solids and structures}, vol.~43, no.~5, pp.~1047--1060, 2006.

\bibitem{zhou2004dynamic}
F.~Zhou and J.-F. Molinari, ``Dynamic crack propagation with cohesive elements:
  a methodology to address mesh dependency,'' {\em International Journal for
  Numerical Methods in Engineering}, vol.~59, no.~1, pp.~1--24, 2004.

\bibitem{kuhn2010continuum}
C.~Kuhn and R.~M{\"u}ller, ``A continuum phase field model for fracture,'' {\em
  Engineering Fracture Mechanics}, vol.~77, no.~18, pp.~3625--3634, 2010.

\bibitem{bourdin2008variational}
B.~Bourdin, G.~A. Francfort, and J.-J. Marigo, ``The variational approach to
  fracture,'' {\em Journal of elasticity}, vol.~91, no.~1-3, pp.~5--148, 2008.

\bibitem{del2013variational}
G.~Del~Piero, ``A variational approach to fracture and other inelastic
  phenomena,'' {\em Journal of Elasticity}, vol.~112, no.~1, pp.~3--77, 2013.

\bibitem{lee2016pressure}
S.~Lee, M.~F. Wheeler, and T.~Wick, ``Pressure and fluid-driven fracture
  propagation in porous media using an adaptive finite element phase field
  model,'' {\em Computer Methods in Applied Mechanics and Engineering},
  vol.~305, pp.~111--132, 2016.

\bibitem{schrefler2006adaptive}
B.~A. Schrefler, S.~Secchi, and L.~Simoni, ``On adaptive refinement techniques
  in multi-field problems including cohesive fracture,'' {\em Computer methods
  in applied mechanics and engineering}, vol.~195, no.~4-6, pp.~444--461, 2006.

\bibitem{khoei2008modeling}
A.~Khoei, H.~Azadi, and H.~Moslemi, ``Modeling of crack propagation via an
  automatic adaptive mesh refinement based on modified superconvergent patch
  recovery technique,'' {\em Engineering Fracture Mechanics}, vol.~75, no.~10,
  pp.~2921--2945, 2008.

\bibitem{tada2000analysis}
H.~Tada, P.~Paris, and G.~Irwin, ``The analysis of cracks handbook,'' {\em New
  York: ASME Press}, vol.~2, p.~1, 2000.

\bibitem{lacazette1990application}
A.~Lacazette, ``Application of linear elastic fracture mechanics to the
  quantitative evaluation of fluid-inclusion decrepitation,'' {\em Geology},
  vol.~18, no.~8, pp.~782--785, 1990.

\bibitem{moelans2008introduction}
N.~Moelans, B.~Blanpain, and P.~Wollants, ``An introduction to phase-field
  modeling of microstructure evolution,'' {\em Calphad}, vol.~32, no.~2,
  pp.~268--294, 2008.

\bibitem{moelans2011quantitative}
N.~Moelans, ``A quantitative and thermodynamically consistent phase-field
  interpolation function for multi-phase systems,'' {\em Acta Materialia},
  vol.~59, no.~3, pp.~1077--1086, 2011.

\bibitem{moelans2008phase}
N.~Moelans, ``A phase-field model for multi-component and multi-phase
  systems,'' {\em Archives of Metallurgy and Materials}, vol.~53, no.~4,
  pp.~1149--1156, 2008.

\bibitem{ribot2019new}
J.~M.~G. Ribot, V.~Agrawal, and B.~S. Runnels, ``A new approach for phase field
  modeling of grain boundaries with strongly nonconvex energy,'' {\em Modelling
  and Simulation in Materials Science and Engineering}, 2019.

\bibitem{zeghadi2007ensemble}
A.~Zeghadi, F.~N'guyen, S.~Forest, A.-F. Gourgues, and O.~Bouaziz, ``Ensemble
  averaging stress--strain fields in polycrystalline aggregates with a
  constrained surface microstructure--part 1: Anisotropic elastic behaviour,''
  {\em Philosophical Magazine}, vol.~87, no.~8-9, pp.~1401--1424, 2007.

\bibitem{brenner2009elastic}
R.~Brenner, R.~Lebensohn, and O.~Castelnau, ``Elastic anisotropy and yield
  surface estimates of polycrystals,'' {\em International Journal of Solids and
  Structures}, vol.~46, no.~16, pp.~3018--3026, 2009.

\bibitem{vidyasagar2017predicting}
A.~Vidyasagar, W.~L. Tan, and D.~M. Kochmann, ``Predicting the effective
  response of bulk polycrystalline ferroelectric ceramics via improved spectral
  phase field methods,'' {\em Journal of the Mechanics and Physics of Solids},
  vol.~106, pp.~133--151, 2017.

\bibitem{towns2014xsede}
J.~Towns, T.~Cockerill, M.~Dahan, I.~Foster, K.~Gaither, A.~Grimshaw,
  V.~Hazlewood, S.~Lathrop, D.~Lifka, G.~D. Peterson, {\em et~al.}, ``Xsede:
  accelerating scientific discovery,'' {\em Computing in Science \&
  Engineering}, vol.~16, no.~5, pp.~62--74, 2014.

\end{thebibliography}

\appendix
\numberwithin{figure}{section}
\section{Coarse/fine stencil}\label{sec:coarsefinestencil}

\begin{figure}[h]
  \centering\includegraphics[width=0.5\linewidth]{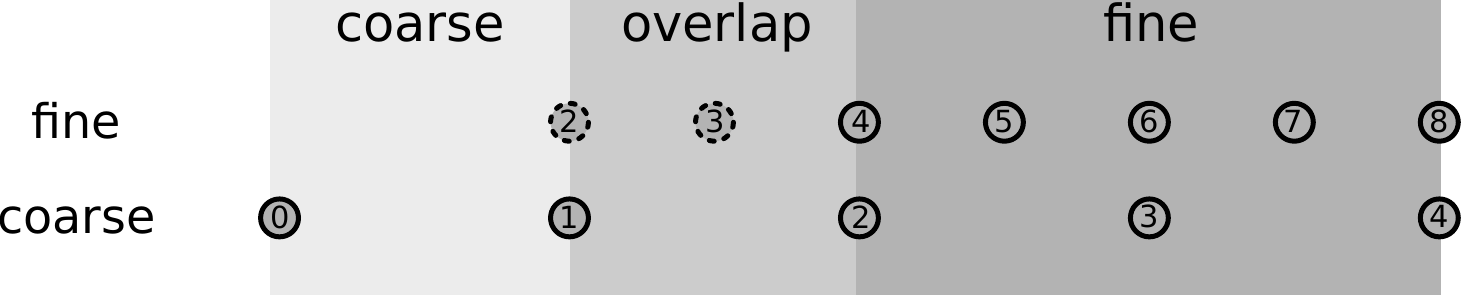}
  \caption{1D example of the grid at the C/F boundary}
  \label{fig:coarsefine}
\end{figure}

It is instructive to find the equivalent numerical stencil corresponding to the reflux-free restriction operator.
We begin with a one-dimensional example.
Let $\phi_{f,4}$ be the fine node on the C/F boundary and $\phi_{c,2}$ be the coarse node on the C/F boundary as shown in Figure~\ref{fig:coarsefine}. 
Let $\Delta x_c$ and $\Delta x_f$ be the spacings on the coarse and fine levels, respectively.
The dashed lines indicate that the nodes are ghost nodes, and the ``overlap'' region is the coarse region over which the ghost nodes are located. 
As a simple yet representative example, let the field be single-component and let the differential operator be a simple Laplacian:
\begin{align}
  D[\phi] = \frac{d^2\phi}{dx^2}.
\end{align}
Then, the operator acting on $\phi_{f,4}$ (i.e. the fine solution) is
\begin{align}\label{eq:1dfineoperator}
  D[\phi_{f,4}] = \frac{\phi_{f,5} - 2\phi_{f,4} + \phi_{f,3}}{\Delta x_f^2}
\end{align}
Since $\phi_{f,3}$ is restricted from the coarse level, then
\begin{align}\label{eq:restrictedfromcoarse}
  \phi_{f,3} = \frac{1}{2}(\phi_{c,1}+\phi_{c,2}) = \frac{1}{2}(\phi_{c,1}+\phi_{f,4}).
\end{align}
Substituting (\ref{eq:restrictedfromcoarse}) into (\ref{eq:1dfineoperator}) yields
\begin{align}
  D[\phi_{f,4}] &= \frac{\phi_{f,5} - 2\phi_{f,4} + \frac{1}{2}(\phi_{c,1}+\phi_{f,4})}{\Delta x_f^2} \\
  &= \frac{\phi_{f,5} - \frac{3}{2}\phi_{f,4} + \frac{1}{2}\phi_{c,1}}{\Delta x_f^2},
\end{align}
which may be alternatively written
\begin{align}
  D[\phi_{f,4}] &= 
                  \frac{\phi_{f,5} - \phi_{f,4} }{\Delta x_f^2} 
                  -
                  2\frac{\phi_{c,2}-\phi_{c,1}}{\Delta x_c^2} ,
\end{align}
corresponding to the Taylor series expansion of the operator at those points.

\begin{figure}[h]
  \centering\includegraphics[width=0.5\linewidth]{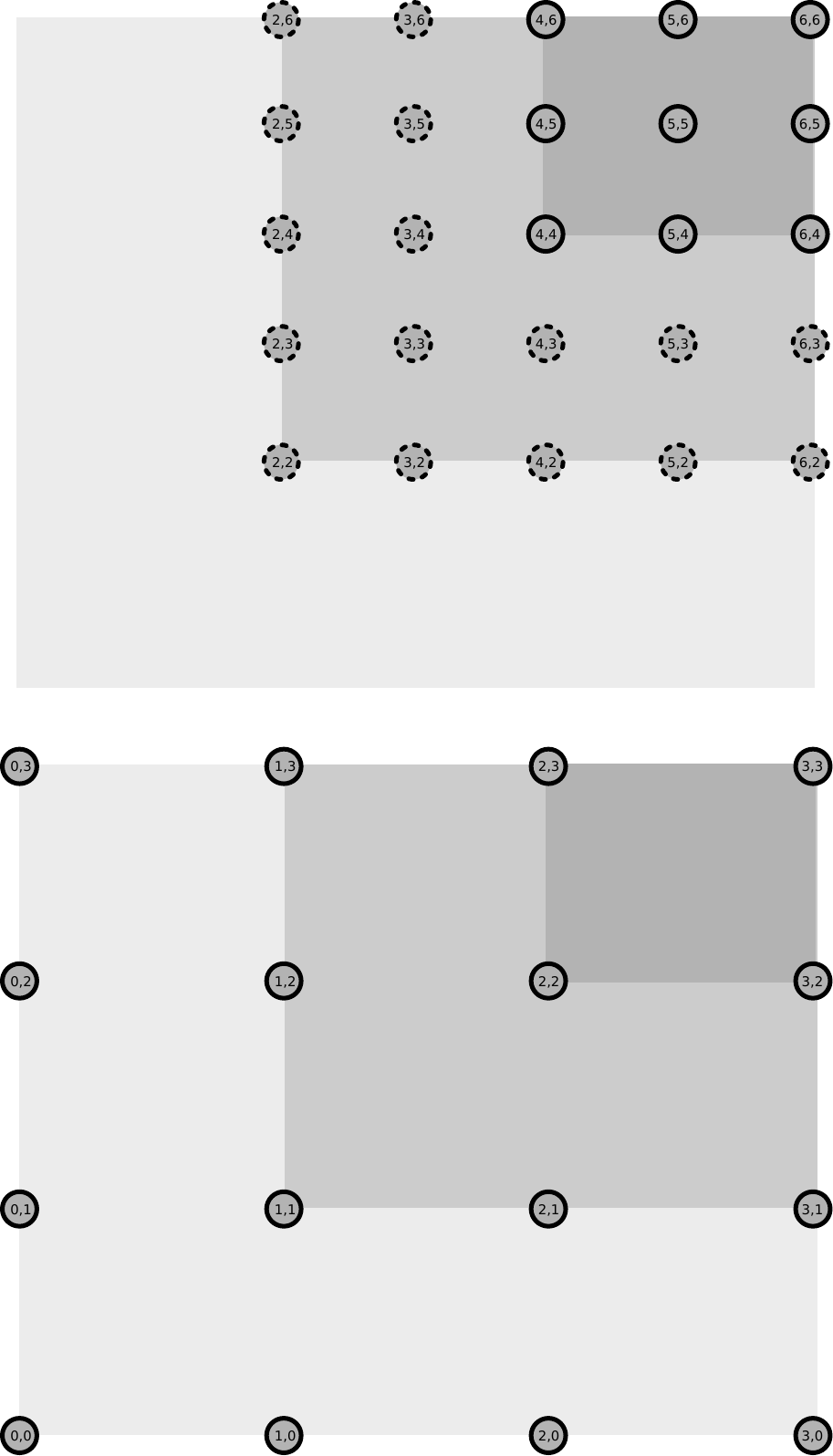}
  \caption{
    Grid at a coarse/fine boundary for the fine grid (top) and coarse grid (bottom)
    The light gray region is coarse-only.
    The dark gray region is fine-only.
    The medium gray region is the overlap region over which the ghost nodes are defined.
  }
  \label{fig:coarsefine2d}
\end{figure}

Now, let us revise to consider the effective stencil defined for a corner node in 2D.
Let the operator be the Laplacian in 2D
\begin{align}
  D[\phi] = \frac{\partial^2\phi}{\partial x^2} + \frac{\partial^2\phi}{\partial y^2}
\end{align}
and let the stencil be defined as illustrated in Figure~\ref{fig:coarsefine2d}.
The operator acting at fine node $\phi_{f,4,4}$ is
\begin{align}\label{eq:2dfinestencil}
  D[\phi_{f,4,4}] =& 
                     \frac{\phi_{f,5,4} - 2\phi_{f,4,4} + \phi_{f,3,4}}{\Delta x_f^2} +\notag\\
                     & \frac{\phi_{f,4,5} - 2\phi_{f,4,4} + \phi_{f,4,3}}{\Delta y_f^2} 
\end{align}
Again realizing that $\phi_{f,4,3}$ and $\phi_{f,3,4}$ are restricted from the coarse grid, we have
\begin{gather}\label{eq:2dfinerestriction}
  \phi_{f,3,4} = \frac{1}{2}(\phi_{c,1,2}+\phi_{c,2,2}) \\
  \phi_{f,4,3} = \frac{1}{2}(\phi_{c,2,1}+\phi_{c,2,2}).
\end{gather}
Substituting (\ref{eq:2dfinerestriction}) into (\ref{eq:2dfinestencil}) we have
\begin{align}
  D[\phi_{f,4,4}] =& 
                     \frac{\phi_{f,5,4} - 2\phi_{f,4,4} + \frac{1}{2}(\phi_{c,1,2}+\phi_{c,2,2})}{\Delta x_f^2} +\notag\\
                     & \frac{\phi_{f,4,5} - 2\phi_{f,4,4} +\frac{1}{2}(\phi_{c,2,1}+\phi_{c,2,2})}{\Delta y_f^2},
\end{align}
which can be written alternatively as either
\begin{align}
  D[\phi_{f,4,4}] =& 
                     \frac{\phi_{f,5,4} - \frac{3}{2}\phi_{f,4,4} + \frac{1}{2}\phi_{c,1,2}}{\Delta x_f^2} +\notag\\
                     & \frac{\phi_{f,4,5} - \frac{3}{2}\phi_{f,4,4} + \frac{1}{2}\phi_{c,2,1}}{\Delta y_f^2} 
\end{align}
or
\begin{align}
  D[\phi_{f,4,4}] =& 
                     \frac{\phi_{f,5,4} - \phi_{f,4,4}}{\Delta x_f^2} -
                     2\frac{\phi_{c,2,2}-\phi_{c,1,2}}{\Delta x_c^2} +\notag\\
                     & \frac{\phi_{f,4,5} - \phi_{f,4,4}}{\Delta y_f^2} - 
                     2\frac{\phi_{c,2,2} - \phi_{c,2,1}}{\Delta y_c^2}.
\end{align}
A similar procedure follows for other nodes along the coarse/fine boundary.

\end{document}